\documentclass[fleqn,usenatbib]{mnras}

\usepackage{newtxtext,newtxmath}

\usepackage[T1]{fontenc}
\usepackage{xcolor}

\DeclareRobustCommand{\VAN}[3]{#2}
\let\VANthebibliography\thebibliography
\def\thebibliography{\DeclareRobustCommand{\VAN}[3]{##3}\VANthebibliography}

\usepackage{graphicx}	
\usepackage{amsmath}	
\usepackage{orcidlink}  
\usepackage{hyperref}   

\newcommand{\NII}{N~{\sc ii}}
\newcommand{\OIII}{O~{\sc iii}}
\newcommand{\HII}{H~{\sc ii}}

\newcommand{\HeII}{He~{\sc ii}}
\newcommand{\HeI}{He~{\sc i}}
\newcommand{\SII}{S~{\sc ii}}
\newcommand{\CaII}{Ca~{\sc ii}}
\newcommand{\NaI}{Na~{\sc i}}
\newcommand{\MgI}{Mg~{\sc i}}
\newcommand{\MgII}{Mg~{\sc ii}}
\newcommand{\FeII}{Fe~{\sc ii}}

\title[ASASSN-23bd: The Lowest Redshift and Least Luminous Optically-Selected TDE To Date]{Discovery and Follow-up of ASASSN-23bd (AT~2023clx): The Lowest Redshift and Luminosity Optically-Selected Tidal Disruption Event}

\author[W. B. Hoogendam et al. 2024]{W. B. Hoogendam\orcidlink{0000-0003-3953-9532},$^{1}$\thanks{E-mail: willemh@hawaii.edu}\thanks{NSF Fellow}
J. T. Hinkle\orcidlink{0000-0001-9668-2920}$^{1}$\thanks{FINESST FI},
B. J. Shappee\orcidlink{0000-0003-4631-1149}$^{1}$,
K. Auchettl\orcidlink{0000-0002-4449-9152}$^{2,3}$,
C.~S.~Kochanek\orcidlink{0000-0001-6017-2961}$^{4,5}$,
K.~ Z.~Stanek$^{4,5}$,
\newauthor
W.~P.~Maksym\orcidlink{0000-0002-2203-7889}$^{6}$,
M.~A.~Tucker\orcidlink{0000-0002-2471-8442}$^{4,5,7}$\thanks{CCAPP Fellow}, 
M.~E.~Huber\orcidlink{0000-0003-1059-9603}$^1$,
N. Morrell\orcidlink{0000-0003-2535-3091}$^8$,
C. R. Burns\orcidlink{0000-0003-4625-6629}$^9$,
D. Hey\orcidlink{0000-0003-3244-5357}$^1$,
\newauthor
T.~W.-S.~Holoien\orcidlink{0000-0001-9206-3460}$^9$,
J.~L.~Prieto$^{10,11}$
M. Stritzinger \orcidlink{0000-0002-5571-1833}$^{12}$,
A.~Do\orcidlink{0000-0003-3429-7845}$^1$, 
A. Polin\orcidlink{0000-0002-1633-6495}$^{9,13,14}$,
C.~Ashall\orcidlink{0000-0002-5221-7557}$^{15}$,
P. J. Brown\orcidlink{0000-0001-6272-5507}$^{16,17}$,
\newauthor
J. M. DerKacy\orcidlink{0000-0002-7566-6080}$^{15}$,
L. Ferrari $^{18,19}$,
L. Galbany\orcidlink{0000-0002-1296-6887}$^{20,21}$,
E. Y. Hsiao\orcidlink{0000-0003-1039-2928}$^{22}$
S. Kumar\orcidlink{0000-0001-8367-7591}$^{23,22}$
J.~Lu\orcidlink{[0000-0002-3900-1452]}$^{24}$, \&
C. P. Stevens\orcidlink{0000-0003-0763-6004}$^{15}$
\\
$^{1}$Institute for Astronomy, University of Hawaii, 2680 Woodlawn Drive, Honolulu, HI 96822, USA\\
$^{2}$School of Physics, The University of Melbourne, Parkville, VIC 3010, Australia\\
$^{3}$Department of Astronomy and Astrophysics, University of California, Santa Cruz, CA 95064, USA\\
$^{4}$Department of Astronomy, The Ohio State University, 140 West 18th Avenue, Columbus, OH 43210, USA\\
$^{5}$Center for Cosmology and AstroParticle Physics, The Ohio State University, 191 W.\ Woodruff Ave., Columbus, OH 43210, USA \\
$^{6}$NASA Marshall Space Flight Center, Huntsville, AL 35812, USA \\
$^{7}$Department of Physics, The Ohio State University, 191 West Woodruff Ave, Columbus, OH, USA\\
$^{8}$Carnegie Observatories, Las Campanas Observatory, Casilla 601, La Serena, Chile\\
$^{9}$The Observatories of the Carnegie Institution for Science, 813 Santa Barbara St., Pasadena, CA 91101, USA\\
$^{10}$ Núcleo de Astronomía de la Facultad de Ingeniería y Ciencias, Universidad Diego Portales, Av. Ejército 441, Santiago, Chile\\
$^{11}$ Millennium Institute of Astrophysics, Santiago, Chile\\
$^{12}$Department of Physics and Astronomy, Aarhus University, Ny Munkegade 120, DK-8000 Aarhus C, Denmark\\
$^{13}$TAPIR, Walter Burke Institute for Theoretical Physics, 350-17, Caltech, Pasadena, CA 91125, USA\\
${^14}$Department of Physics and Astronomy, Purdue University, 525 Northwestern Avenue, West Lafayette, IN 47907, USA\\
$^{15}$Department of Physics, Virginia Tech, Blacksburg, VA 24061, USA\\
$^{16}$George P. and Cynthia Woods Mitchell Institute for Fundamental Physics and Astronomy\\ 
$^{17}$Department of Physics and Astronomy, Texas A\&M University, College Station, TX 77843, USA \\
$^{18}$Institute of Space Sciences (ICE-CSIC), Campus UAB, Carrer de Can Magrans, s/n, E-08193 Barcelona, Spain.\\
$^{19}$Institut d’Estudis Espacials de Catalunya (IEEC), E-08034 Barcelona, Spain\\
$^{20}$Institute of Space Sciences (ICE-CSIC), Campus UAB, Carrer de Can Magrans, s/n, E-08193 Barcelona, Spain.\\
$^{21}$Institut d’Estudis Espacials de Catalunya (IEEC), E-08034 Barcelona, Spain\\
$^{22}$Department of Physics, Florida State University, 77 Chieftan Way, Tallahassee, FL 32306, USA\\
$^{23}$Department of Astronomy, University of Virginia, 530 McCormick Rd, Charlottesville, VA 22904, USA\\
$^{24}$Department of Physics and Astronomy, Michigan State University, East Lansing, MI 48824, USA\\
}

\date{2024 April 22. Received 2024 April 22; in original form 2024 January 10}

\pubyear{2024}

\begin{document}
\label{firstpage}
\pagerange{\pageref{firstpage}--\pageref{lastpage}}
\maketitle

\begin{abstract}
We report the All-Sky Automated Survey for SuperNovae discovery of the tidal disruption event (TDE) ASASSN-23bd (AT~2023clx) in NGC 3799, a LINER galaxy with no evidence of strong AGN activity over the past decade. With a redshift of $z = 0.01107$ and a peak UV/optical luminosity of $(5.4\pm0.4)\times10^{42}$ erg s$^{-1}$, ASASSN-23bd is the lowest-redshift and least-luminous TDE discovered to date. Spectroscopically, ASASSN-23bd shows  H$\alpha$ and \HeI\ emission throughout its spectral time series, there are no coronal lines in its NIR spectrum, and the UV spectrum shows nitrogen lines without the strong carbon and magnesium lines typically seen for AGN. Fits to the rising ASAS-SN light curve show that ASASSN-23bd started to brighten on MJD 59988$^{+1}_{-1}$, $\sim$9 days before discovery, with a nearly linear rise in flux, peaking in the $g$ band on MJD $60000^{+3}_{-3}$. Scaling relations and TDE light curve modelling find a black hole mass of $\sim$10$^6$ $M_\odot$, which is on the lower end of supermassive black hole masses. 
ASASSN-23bd is a dim X-ray source, with an upper limit of $L_{0.3-10\,\mathrm{keV}} < 1.0\times10^{40}$ erg s$^{-1}$ from stacking all \emph{Swift} observations prior to MJD 60061, but with soft ($\sim 0.1$~keV) thermal emission with a luminosity of $L_{0.3-2 \,\mathrm{keV}}\sim4\times10^{39}$ erg s$^{-1}$ in \emph{XMM-Newton} observations on MJD 60095.
The rapid $(t < 15$~days) light curve rise, low UV/optical luminosity, and a luminosity decline over 40 days of $\Delta L_{40}\approx-0.7$~dex make ASASSN-23bd one of the dimmest TDEs to date and a member of the growing ``Low Luminosity and Fast'' class of TDEs.  
\end{abstract}

\begin{keywords}
transients: tidal disruption events -- black hole physics -- accretion, accretion discs
\end{keywords}



\section{Introduction}\label{sec:intro}

Tidal disruption events (TDEs) are the result of the partial or total disruption of a star passing near or within the tidal radius of a supermassive black hole (SMBH), leading to a luminous flare from accretion onto the SMBH (e.g., \citealt{rees88,evans89,phinney89,ulmer99,komossa15,stone19}). Emission from TDEs occurs over a broad range of wavelengths, including the hard X-ray (e.g., \citealt{bloom11,burrows11,cenko12b,pasham15}), soft X-ray (e.g., \citealt{bade96,grupe99,komossa99b,auchettl17}), ultraviolet (UV, e.g., \citealt{stern04,gezari06,gezari08,gezari09}), and optical bands (e.g., \citealt{vanvelzen11,cenko12a,gezari12b,arcavi14,chornock14,holoien14b,vinko15,holoien16b,holoien16c,brown18,holoien19a,holoien19b}). While active galactic nuclei (AGNs) probe actively accreting SMBHs, TDEs offer a rare opportunity to examine the activation of otherwise dormant SMBHs (e.g., \citealt{lodato11,guillochon15,shiokawa15,metzger16}). TDEs also provide a laboratory for studying shock physics (e.g., \citealt{lodato09}), jet formation (e.g., \citealt{farrar14,wang16,biehl18}), and the local environment and growth of SMBHs (e.g., \citealt{auchettl18,jiang21b,mockler22}). 

However, the physics that underlies TDEs is complex. Possible parameters include the age, mass, composition, and rotation of the disrupted star \citep{kochanek16a,gallegos-garcia18,law-smith19,mockler19,golightly19}, the spin and mass of the SMBH \citep{ulmer99,graham01,mockler19,gafton19}, the stellar impact parameter \citep{guillochon13,guillochon15,gafton19}, the accretion fraction and viewing angle \citep{kochanek94,lodato11,dai15,guillochon15,shiokawa15,metzger16,dai18,coughlin19}.

Despite all these parameters, optical and UV spectral energy distributions (SEDs) of TDEs are well modeled by blackbodies (e.g., \citealt{gezari12b,holoien14a,holoien16b,holoien16c,brown16a,hung17,holoien18a,holoien19b,leloudas19,holoien20,vanvelzen21a,hinkle21b}). 

Several methods have been proposed to measure black hole mass from the TDE light curve (e.g., \citealt{ulmer99,graham01,gafton19,mockler19,ryu20,Wen20,Ramsden22,Mummery23,Sarin24}), these have been able to reproduce the known SMBH scaling relations with galaxy properties with varying success (e.g., \citealt{hammerstein23}).
Finally, TDEs also show a \citet{phillips93}-esque relationship between peak bolometric luminosity and decline rate \citep{hinkle20a,hinkle21b,hammerstein23}. 

From an optical spectroscopic perspective, the presence, strength and width of the observed lines vary (e.g., \citealt{arcavi14,hung17,leloudas19,wevers19,holoien20,vanvelzen20b,Charalampopoulos22,Nicholl22}). Generally, optical emission lines are observed from hydrogen and/or helium, occasionally complemented by oxygen lines from Bowen fluorescence (e.g., \citealt{leloudas19,vanvelzen20b}). Several potential theoretical explanations have been proposed to explain the variations in the spectra of TDEs, including photoionization physics (e.g., \citealt{gaskell14,guillochon14,roth16,kara18,leloudas19}), stellar composition differences (e.g., \citealt{kochanek16a,law-smith20, mockler22}), and He star progenitors (e.g., \citealt{gezari12b,strubbe15}). While TDEs mostly have broad, Gaussian line profiles, some TDEs show line profiles with either strong, narrow line profiles (e.g., \citealt{holoien20,vanvelzen20b}) or double peaks (e.g., \citealt{holoien19a,hung20}), which may be explained by the viewing geometry (e.g., \citealt{holoien19a,short20,Hung21}). Finally, TDEs separate into diversely behaving spectral classes \citep{leloudas19,vanvelzen21a,hammerstein23}, which has potential physical explanations that include partial vs. complete disruption \citep{Nicholl22} and ionization from a shrinking photospheric radius \citep{Charalampopoulos22}. However, a complete satisfactory explanation for the spectral classes is still needed. A larger sample of well-observed TDEs should advance our understanding of the underlying nature of TDEs and the SMBHs that cause them.

Fortunately, with many transient surveys currently operating such as the All-Sky Automated Survey for Supernovae (ASAS-SN; \citealt{shappee14,kochanek17}), the Asteroid Terrestrial Impact Last Alert System (ATLAS; \citealt{tonry18}), the Panoramic Survey Telescope and Rapid Response System (Pan-STARRS; \citealt{chambers16, Jones21}), and the Zwicky Transient Facility (ZTF; \citealt{bellm19}), more nuclear transients, including TDEs, are discovered each year, and a growing number are now found prior to peak brightness (e.g., \citealt{holoien19a,holoien19b,leloudas19,vanvelzen19,wevers19,holoien20,vanvelzen20b}). The rising light curve and pre-maximum spectra provide important information about TDEs, such as the progression from a disrupted star to an accretion flow or an initial X-ray pulse \citep{Carter83, Brassart08}. The early detection of TDEs is especially important for a growing subclass of Faint and Fast (FaF) TDEs\footnote{In this context, faint refers to intrinsically faint, not observationally faint.}. As their name suggests, the peak luminosity of a FaF TDE tends to be an order of magnitude less luminous combined with a rise and decline about twice as fast as a ``normal'' TDE. Examples of FaF TDEs include iPTF16fnl \citep{blagorodnova17,brown18}, ATLAS18mlw \citep{hinkle23a}, AT~2019qiz \citep{nicholl20}, AT~2020neh \citep{Angus22}, and AT~2020wey \citep{Charalampopoulos23}. While the observed sample is presently small, FaF TDEs may have a higher intrinsic rate than ``normal'' TDEs \citep{Charalampopoulos23} but a lower discovery rate due to their lower luminosity. This is consistent with the findings of \citet{vanvelzen18} and \citet{Yao23} that TDEs have a steep luminosity function.

In addition to TDEs, transient sky surveys have uncovered additional classes of nuclear transients, including a number of ambiguous transients. For example, the differences between a TDE and single bright AGN flare are poorly understood (see \citealt{Zabludoff21} for a review), or, alternatively, changing-look AGN may originate from accretion rate changes caused by a TDE (e.g., \citealp{Chan19, vanvelzen20b, Zhang21, li22, Ricci23}). The potential for mistaken identification is particularly a problem for galaxies that already host AGN activity. In particular, another growing class of nuclear transients is ambiguous nuclear transients (ANTs), which exhibit observational characteristics seen in both AGNs and TDEs (e.g., \citealp{trakhtenbrot19b} \citealp{neustadt20}, \citealp{hinkle22a}, \citealp{li22}, \citealp{Holoien22}). While one may posit ANTs are merely TDEs in AGN hosts, some ANTs occur in host galaxies that lack active accretion (e.g., \citealt{malyali21}, \citealt{hinkle22a}, and \citealt{Holoien22}).

Here, we present the discovery and follow-up observations of the nuclear transient ASASSN-23bd. We present the data in Section \ref{sec:data} and analysis in Section \ref{sec:analysis}. Section \ref{sec:discussion} compares ASASSN-23bd with other FaF TDEs, and Section \ref{sec:Summary} summarises the results.

\section{Data} \label{sec:data}
\subsection{Initial Discovery and Classification}
ASAS-SN discovered ASASSN-23bd (a.k.a. AT~2023clx\footnote{\href{https://www.wis-tns.org/object/2023clx}{https://www.wis-tns.org/object/2023clx}}) on MJD 59997.2 at ($\alpha,\delta$) $=$ (11:40:09.397 $+$15:19:38.54) in NGC 3799 using the Cassius unit in Chile \citep{ASASSN-23bd_disc}. The ASAS-SN discovery $g$-band magnitude was 16.3, with a last non-detection on MJD 59988.3 at a limiting magnitude of $g \ge 17.9$. \citet{ASASSN-23bd_class} spectroscopically classified ASASSN-23bd as a TDE on MJD 60001.7. 

The redshift of NGC 3799 is $z=0.01107$ \citep{SDSS_DR_13}, making ASASSN-23bd the lowest redshift TDE to date. \citet{Theureau07} derive distance moduli of $33.6\pm0.4$ mag ($\sim$51 Mpc; 21-cm line) and $35.6\pm0.5$ mag ($\sim$134 Mpc; H-band \citealt{Tully77} relationship). Assuming $H_0 = 73$~km~sec$^{-1}$~Mpc$^{-1}$ \citep{Burns18, Riess22, Galbany23}, $\Omega_{m}= 0.3$, and $\Omega_{\Lambda} = 0.7$, the distance modulus to NGC 3799 is $33.50\pm0.15$ mag ($50.1\pm3.5$ Mpc) in the cosmic microwave background frame. We adopt this distance.  

\subsection{Survey Data}

\begin{table*}
\centering
\caption{Host-subtracted stacked ATLAS and ASAS-SN photometry of ASASSN-23bd. Uncertainties of 99.999 denote 3$\sigma$ upper limits. Upper and lower uncertainties on the date of observation are derived during the stacking procedure to span the distance between the earliest and latest photometric epochs combined to result in that datum point. The full table will be available in the online journal.}\label{tab:surveys_phot}
\begin{tabular}{cccccc}
\hline
JD & Filter & Magnitude & Uncertainty &  Flux [mJy] & Uncertainty\\ 
\hline
2459967.03$^{0.01}_{0.01}$ & $o$ & 20.297 & 99.999 & $-$0.076 & 0.006	 \\
2459970.14$^{0.02}_{0.01}$ & $o$ & 19.862 & 99.999 & $-$0.068 & 0.008	 \\
2459979.86$^{0.03}_{0.04}$ & $o$ & 19.325 & 99.999 & $-$0.046 & 0.014	 \\
\hline
\end{tabular}
\end{table*}

\subsubsection{ASAS-SN Light Curve}
ASAS-SN is a fully robotic survey with 20 14-cm telescopes distributed on five mounts at four sites, providing comprehensive all-sky monitoring with a cadence of $\sim$20 hours (in good conditions) designed to detect nearby supernovae. The five ASAS-SN units are located at the Haleakala Observatory, the South African Astrophysical Observatory, the McDonald Observatory, and two at the Cerro Tololo Inter-American Observatory. The typical ASAS-SN observing strategy is to obtain 3 dithered images at each pointing. The ASAS-SN survey began observations in late 2011 with the $V$ band. In 2017, ASAS-SN added 12 telescopes using the $g$ band, and the original eight telescopes were switched from the $V$ band to the $g$ band in 2018.  

ASAS-SN observed the location of ASASSN-23bd 3441 times before discovery, and all images were reduced using the standard ASAS-SN pipeline based on the ISIS image subtraction package \citep{alard98,alard00}. We used images taken before MJD 59800 to construct the reference image, removing any images with a large FWHM ($\ge1.7$ pixels), a 3-sigma depth that was too shallow ($g < 17.0$ mag), or that showed signs of cirrus or clouds.

Similar to the standard ASAS-SN Sky Patrol photometry \citep{kochanek17, Hart23}, we use the IRAF \textsc{apphot} package with a 2-pixel ($16$ arcsecond) radius to perform aperture photometry on each subtracted image, generating a differential light curve. The photometry is calibrated using the AAVSO Photometric All-Sky Survey \citep{Henden15}. 
We then stack the individual ASAS-SN images on different time scales for different parts of the light curve.  First, when looking for pre-discovery variability we stack the dithered images together.  Next, from 40 days before to 25 days after discovery, we stack in 25-hour bins to cover the rapid rise and decline. Lastly, for the declining light curve, we stack in 100-hour bins to better follow the relatively slow fading. The ASAS-SN photometric observations are compiled in Table \ref{tab:surveys_phot}.

\subsubsection{ATLAS Light Curve}
The ATLAS survey primarily focuses on detecting small asteroids that have a chance of terrestrial collision using two filters: a cyan ($c$; 420-650 nm) filter and an orange ($o$; 560-820nm) filter \citep{tonry18}. ATLAS uses 0.5 m Wright-Schmidt telescopes in Hawaii, Chile, and South Africa to obtain four 30-second exposures in an hour-long window for 200–250 fields per night, covering approximately a quarter of the sky \citep{smith20}. Data were retrieved from the ATLAS Transient Science Server \citep{smith20}. The ATLAS light curve contains 2,527 images, with the earliest on MJD 57400.6. We combined the four nightly ATLAS images to derive a light curve with 630 epochs. The initial rise was observed in the ATLAS $o$ band, whereas the peak was observed in the ATLAS $c$ band. Data taken on the same night were stacked using a weighted average, excluding data affected by clouds. The ATLAS photometric observations are also compiled in Table \ref{tab:surveys_phot}.

\subsubsection{TESS Data}
NGC 3799 was observed by the Transiting Exoplanet Survey Satellite (TESS; \citealt{ricker15}) in sectors 22, 45, 46, and 49. Unfortunately, these sectors all occurred before the discovery of ASASSN-23bd. We reduced the TESS data in a similar manner to the ASAS-SN data following the processes detailed in \citet{vallely19}, \citet{vallely21}, and \citet{fausnaugh21}. We used the ISIS package \citep{alard98,Alard2000} to image subtract the full-frame TESS images. 
Median filters were used to remove artifacts like CCD straps. We produced light curves from these subtracted images with reference images constructed on a per-sector basis using the first 100 good-quality full-frame images without compromised pointing, significant scattered light, or data quality flags. For the extended mission with its shorter integration times, we use the first 300 images that meet these criteria. 

\subsubsection{ZTF Light Curve}
ZTF uses the Samuel Oschin 48-in Schmidt telescope at the Palomar Observatory and a camera with a 47 deg$^2$ field of view to obtain images as deep as 20.5 $r$-band mag in 30-second exposures. ZTF observed the field containing ASASSN-23bd starting on MJD 58202.3. We use the ZTF $g$- and $r$-band light curves between MJD 58202.3 and MJD 60090 obtained through the ZTF forced photometry service\footnote{\href{https://irsa.ipac.caltech.edu/Missions/ztf.html.}{https://irsa.ipac.caltech.edu/Missions/ztf.html.}}. These light curves have 673 $g$- and $r$-band epochs constructed from 1,726 images. Following the ASAS-SN discovery announcement, ZTF reported a detection at $g=19.28$ mag on MJD $\sim$59980 to TNS; however, our analysis of the ZTF forced photometry does not show a detection at this epoch. All the ZTF detections are after the peak of ASASSN-23bd. Since higher-cadence data are available from the Swope Telescope, as discussed below, we only include the ZTF data in our analysis of previous AGN variability.

\subsection{Follow-up Observations}

\begin{table}
\centering
\caption{Host-subtracted photometry of ASASSN-23bd from \emph{Swift} and Swope. The full table will be available in the online journal. The $B$ and $V$ filter keys correspond to Swope filters.}\label{tab:swift-swope}
\begin{tabular}{cccc}
\hline
JD & Filter & Magnitude & Uncertainty \\ 
\hline
2460003.25 & $U$ & 16.42 & 0.06	 \\
2460003.25 & $w1$ & 16.79 & 0.07	 \\
2460003.26 & $m2$ & 16.88 & 0.05	 \\
\hline
\end{tabular}
\end{table}

\subsubsection{Swope Observations}
As part of the Precision Observations of Infant Supernova Explosions (POISE; \citealp{POISE_ATel}) collaboration, we obtained follow-up images from the 1.0-m Henrietta Swope Telescope. These data were taken in the Carnegie Supernova Project (CSP) natural system for which the $BV$ photometry is calibrated using standards from \citet{Landolt07} and the $gri$ photometry is calibrated using standards from \citet{Smith02}. These standards are converted to the CSP system using colour terms from \citet{Krisciunas17} and \citet{Phillips19}. For more information on the CSP filter system, see \citet{Stritzinger11} and references therein. 

The Swope photometry is template-subtracted using Pan-STARRS \citep{chambers16} imaging data. Calibration is done using RefCat2 \citep{tonry18b} magnitudes transformed to the CSP natural system using the colour terms found in \citet{Krisciunas17} and \citet{Phillips19}. Finally, we use the corrections listed on the CSP website\footnote{\href{https://csp.obs.carnegiescience.edu/data/filters}{https://csp.obs.carnegiescience.edu/data/filters}} to convert to the AB system. 

\subsubsection{\emph{Swift} Observations}
The Neil Gehrels Swift Observatory (\emph{Swift}; \citealt{gehrels04}) acquired 19 epochs of data between MJD 60000 and MJD 60061 (PIs: Leloudas, Gomez, Huang, and Wevers). \emph{Swift} simultaneously observed ASASSN-23bd with the UltraViolet and Optical Telescope (UVOT; \citealt{roming05}) and X-Ray Telescope (XRT; \citealt{burrows05}). The \emph{Swift} data through MJD 60061 are included here; later \emph{Swift} observations exist, but these data are too noisy to be useful.

All exposures for each UVOT epoch were combined using the HEASOFT Version 6.31.1 UVOTIMSUM package, and aperture photometry was obtained using the UVOTSOURCE package. An aperture of 5'' is used for both the source and the background. Stacked archival \emph{Swift} host-galaxy images were subtracted in all bands except the $B$ band, which lacks archival \emph{Swift} imaging of the host. We compute Vega-system magnitudes using the \citet{Breeveld11} zero points, which update the \citet{poole08} zero points. A comparison to photometry from the Swift Optical/Ultraviolet Supernova Archive (SOUSA; \citealp{SOUSA}) pipeline yields similar photometry to ours. Finally, the magnitudes are converted to AB magnitudes\footnote{The conversions are found at \\
\href{https://swift.gsfc.nasa.gov/analysis/uvot_digest/zeropts.html}{https://swift.gsfc.nasa.gov/analysis/uvot\_digest/zeropts.html} }. 

Due to the lack of archival $B$-band observations, the large $V$-band uncertainties at late times, and the existence of high quality $B$ and $V$ data from POISE, we elect not to show the \emph{Swift }$B$- or $V$-band photometry nor use the \emph{Swift }$B$-band photometry in our analysis. The \emph{Swift} $V$-band photometry is included in the blackbody/SED fits. The Swope and \emph{Swift} photometry is compiled in Table \ref{tab:swift-swope}.

The \emph{Swift} XRT data was collected in photon-counting mode. Using the most up-to-date calibrations and the standard filters and screenings, the observations were processed using the XRTPIPELINE version 0.13.7. Using a source region with a radius of 47'' centered on the location of ASASSN-23bd and a source-free background region with a radius of 150'' centered at ($\alpha$,$\delta$)=(11:39:56.13,$+$15:22:23.90), no significant X-ray emission associated with the source was found in the individual epochs.

To constrain the X-ray emission, we merged all 19 observations (up to ObsID sw00015897021 on MJD = 60061) using the HEASOFT tool \textsc{xselect} version 2.5b to derive a 3$\sigma$ upper limit of $1\times10^{-3}$ counts/sec for the 0.3-10.0 keV energy range. Assuming an absorbed power law with a Galactic column density of 2.5$\times10^{20}$ cm$^{-2}$ \citep{HI4PI16} and a photon index of 2 at the redshift of the host galaxy, we obtain an absorbed flux limit of $<3.5\times10^{-14}$ erg cm$^{-2}$ s$^{-1}$, which corresponds to an X-ray luminosity limit of $<1\times10^{40}$ erg s$^{-1}$. If we assume an absorbed blackbody model with a temperature of 0.1 keV and a host column density of 1$\times10^{20}$ cm$^{-2}$, as derived in Section \ref{sec:XMM}, we obtain an unabsorbed flux limit of $<2.7\times10^{-14}$ erg cm$^{-2}$ s$^{-1}$, which corresponds to an X-ray luminosity of $<7.9\times10^{39}$ erg s$^{-1}$

\subsubsection{XMM-Newton}\label{sec:XMM}
XMM-Newton targeted ASASSN-23bd through joint time awarded as part of the Hubble GO program 16775 (PI: Maksym). XMM-Newton observed NGC 3799 on 2023 May 31 (MJD 60095; obsid 0892201601) for 15 ks, of which 11.8 ks were useful. ASASSN-23bd was detected at $\sim$10$\sigma$ with $\sim$170 counts (maximum likelihood 185 from the EPIC detection pipeline).
As a first test, we extracted the PN counts from a $r=30\arcsec$ region centred on the source and from a $r=30\arcsec$ sourceless background region near the NW corner of the same PN CCD, avoiding the chip edges. The PN 2.0-10.0 keV excess had $<2\sigma$ significance, with $\sim90\%$ of the photons observed in the 0.1-2 keV band.  Of these, only 1 net count was found between 1.0-2.0 keV.
For a first estimate of spectral properties, we used PIMMS\footnote{\href{https://cxc.harvard.edu/toolkit/pimms.jsp}{https://cxc.harvard.edu/toolkit/pimms.jsp}} to estimate a blackbody temperature assuming a host column density of 1 $\times$ 10$^{20}$ cm$^{-2}$, scaling from our derived host extinction limit ($A_V \le 0.05$; see Section \ref{sec:host}) and Galactic extinction ($A_V = 0.085$; \citealp{schlafly11}) from {\tt colden} \footnote{\href{https://cxc.harvard.edu/toolkit/colden.jsp}{https://cxc.harvard.edu/toolkit/colden.jsp}}. The model blackbody temperature was iteratively increased in steps of 0.01 keV to match the observed hardness ratio such that HR=(H-S)/(H+S)=$-0.38$ where S=0.2-0.5 keV and H=0.5-1.0 keV. This leads to a blackbody temperature estimate of $kT\sim0.09$~keV. 
This leads to an observed flux of $\sim8.8 \times 10^{-15}$ erg s$^{-1}$ cm$^{-2}$ in the Swift 0.3-10 keV band, or an intrinsic $\sim4.2\times10^{39}$ erg s$^{-1}$ for an unabsorbed blackbody in the 0.1-2 keV band.  It is undetected in the hard band ($<1.3\times10^{-14}$ erg s$^{-1}$ cm$^{-2}$ at $3\sigma$ in the 2-10 keV band assuming a $\Gamma=1.7$ power law). 
To check our assumptions, we also extracted a spectrum with {\tt XMM SAS}\footnote{\href{https://www.cosmos.esa.int/web/xmm-newton/sas}{https://www.cosmos.esa.int/web/xmm-newton/sas}} and fit it with XSPec\footnote{\href{https://heasarc.gsfc.nasa.gov/xanadu/xspec/}{https://heasarc.gsfc.nasa.gov/xanadu/xspec/}} using 10-count bins and {\tt lstat} as the minimization statistic. The blackbody fit produces $kT=0.10\pm0.02$, $n_H<1.1\times10^{21}\mathrm{cm}^{-2}$, and $F_{0.3-10\,\mathrm{keV}}=[1.53\pm0.27]\times10^{-14}\mathrm{erg\,s}^{-1}{\mathrm {cm}}^{-2}$ (absorbed) and $[3.17\pm1.85]\times10^{-14}\mathrm{erg\,s}^{-1}{\mathrm{cm}}^{-2}$ (unabsorbed; 90\% confidence), with $\chi^2/\mathrm{DOF}=40.32/25$.
A complete analysis of the XMM-Newton data will be presented in Maksym et al. (2024; in prep.).

\subsubsection{Spectroscopic Observations}

\begin{table}
\centering
\caption{Log of spectroscopic observations.}\label{tab:spec}
\begin{tabular}{ccccc}
\hline
UT Date & MJD & Epoch &  Telescope & Spectrograph \\ 
& [days] & [days] & & \\
\hline
2023-02-26  &   60001.6 &    4.4    &   Seimi   &   KOOLS-IFU   \\
2023-03-04  &   60007.5 &   10.3    &   SS2.3   &   WiFeS       \\
2023-03-05  &   60008.3 &   11.1    &   Baade   &   IMACS       \\
2023-03-14  &   60017.5 &   20.3    &   SS2.3   &   WiFeS       \\
2023-03-15  &   60018.3 &   21.1    &   Baade   &   IMACS       \\
2023-03-17  &   60020.5 &   23.3    &   SS2.3   &   WiFeS       \\
2023-03-19  &   60022.5 &   25.3    &   SS2.3   &   WiFeS       \\
2023-03-20  &   60023.3 &   26.1    &   Baade   &   IMACS       \\
2023-03-20  &   60023.5 &   26.3    &   Keck1   &   LRIS        \\
2023-03-21  &   60024.6 &   27.4    &   SS2.3   &   WiFeS       \\
2023-03-27  &   60030.4 &   33.2    &   UH2.2   &   SNIFS       \\
2023-03-28  &   60031.4 &   34.2    &   UH2.2   &   SNIFS       \\
2023-03-29  &   60032.4 &   35.2    &   UH2.2   &   SNIFS       \\
2023-03-29  &   60032.5 &   35.3    &   SS2.3   &   WiFeS       \\
2023-04-05  &   60039.5 &   42.3    &   SS2.3   &   WiFeS       \\
2023-04-10  &   60044.4 &   47.2    &   UH2.2   &   SNIFS       \\
2023-04-11  &   60045.4 &   48.2    &   UH2.2   &   SNIFS       \\
2023-04-13  &   60047.2 &   50.0    &   Baade   &   IMACS       \\
2023-04-14  &   60048.4 &   51.2    &   SS2.3   &   WiFeS       \\
2023-04-16  &   60050.4 &   53.2    &   UH2.2   &   SNIFS       \\
2023-04-21  &   60055.4 &   58.2    &   UH2.2   &   SNIFS       \\
2023-04-21  &   60055.5 &   58.3    &   SS2.3   &   WiFeS       \\
2023-04-24  &   60058.4 &   61.2    &   UH2.2   &   SNIFS       \\
2023-04-28  &   60062.3 &   65.1    &   UH2.2   &   SNIFS       \\
2023-05-10  &   60074.4 &   77.2    &   UH2.2   &   SNIFS       \\
\hline
2023-03-26 &    60029.4 &   32.2    &   IRTF    &   SpeX     \\
\hline
2023-04-04 &    60038.9 &   41.7    &   HST     &   STIS     \\
\end{tabular}\\
{The epoch phase is relative to the time of discovery on MJD~59997.2 } \\
\end{table}

We acquired 24 spectra of ASASSN-23bd between MJD 60007 and MJD 60075 and also include two public spectra of ASASSN-23bd accessed through the Transient Name Server (TNS) and a public \emph{HST} spectrum. The first TNS spectrum\footnote{Accessed via \href{https://www.wis-tns.org/object/2023clx}{https://www.wis-tns.org/object/2023clx}} was taken on the Seimi telescope using the Kyoto Okayama Optical Low-dispersion Spectrograph Integral Field Unit (KOOLS-IFU; \citealt{Matsubayashi19}). The second TNS spectrum\footnote{Accessed via \href{https://www.wis-tns.org/object/2018meh}{https://www.wis-tns.org/object/2018meh}} was taken on the Keck-I telescope using the Low-Resolution Imaging Spectrometer (LRIS; \citealt{oke95,Rockosi10}) and reported by \citet{TNS_KECK}. 

Our optical spectroscopic observations of ASASSN-23bd are from the POISE \citep{POISE_ATel} and SCAT \citep{Tucker22} collaborations. POISE observations were taken using the Inamori-Magellan Areal Camera and Spectrograph (IMACS; \citealt{Dressler06}) on the 6.5-m Magellan Baade telescope. SCAT observations were taken using the University of Hawaii 2.2m telescope (UH2.2) on Mauna Kea using the Supernova Integral Field Spectrograph (SNIFS; \citealt{lantz04}) and on the Australian National University 2.3m telescope (SS2.3) using the Wide-Field Spectrograph (WiFeS; \citealt{Dopita07,Dopita10}). 

Data reduction for IMACS was performed using standard {\sc{iraf}}\footnote{The Image Reduction and Analysis Facility (IRAF) is distributed by the National Optical Astronomy Observatory, which is operated by the Association of Universities for Research in Astronomy, Inc., under cooperative agreement with the National Science Foundation.} packages and uses the methods described in \citet{hamuy06} and \citet{folatelli13}. SNIFS spectra were reduced using the SCAT pipeline described in \citet{Tucker22}, and WiFeS spectra were reduced using standard procedures implemented in PyWiFeS \citep{Childress14}.

Further spectral data were taken using NASA's InfraRed Telescope Facility (IRTF) with SpeX \citep{Rayner03} as part of IRTF program 2023A060 (PI: Hinkle) and \emph{HST} data were taken with the Space Telescope Imaging Spectrograph (STIS; \citealt{Woodgate98}) as part of GO program 16775 (PI: Maksym). The SpeX data were reduced with telluric corrections from an A0V star using the standard SpeXtool procedures described in \citet{Cushing04}, and the reduced STIS spectrum was obtained through the Mikulski Archive for Space Telescopes. Table \ref{tab:spec} presents a log of the spectroscopic observations of ASASSN-23bd.

\section{Analysis} \label{sec:analysis}
In this section, we analyze the archival host properties of NGC 3799. We search for previous AGN-like variability and compute the general host-galaxy properties. Additionally, we fit the rising light curve of ASASSN-23bd with single- and double-component models and perform blackbody fits to the photometry. Finally, we analyze the optical spectral time series, especially the evolution of the Full-Width at Half-Maximum (FWHM) and luminosity of the $H\alpha$ feature.

\subsection{Archival Observations of Host Galaxy NGC 3799}\label{sec:host}
NGC 3799 is a well-observed galaxy morphologically classified as a SB(s)b:pec \citep{devacouleurs91} galaxy. It is designated as a peculiar galaxy because it is interacting with NGC 3800. There is evidence that galaxies that have undergone recent mergers may be more likely to host a TDE \citep{prieto16, hammerstein21}. NGC 3799 is a Low Ionization Nuclear Emission Line Region (LINER) AGN \citep{Toba14}. The archival SDSS spectrum \citep{york00} shows H$\beta$, Mg $b$ $\lambda$5175, \NaI\ D $\lambda\lambda$5890, 5896, and \CaII\ KHG $\lambda\lambda$3934, 3968, 4308, \CaII\ $\lambda\lambda$8542, 8662, and \MgI\ $\lambda$8807 absorption and [\OIII] $\lambda\lambda$4959,5007, [\NII] $\lambda$6584, [\SII] $\lambda\lambda$6717, 6731, and  H$\alpha$ emission.

\begin{figure*}
    \includegraphics[width=0.49\textwidth]{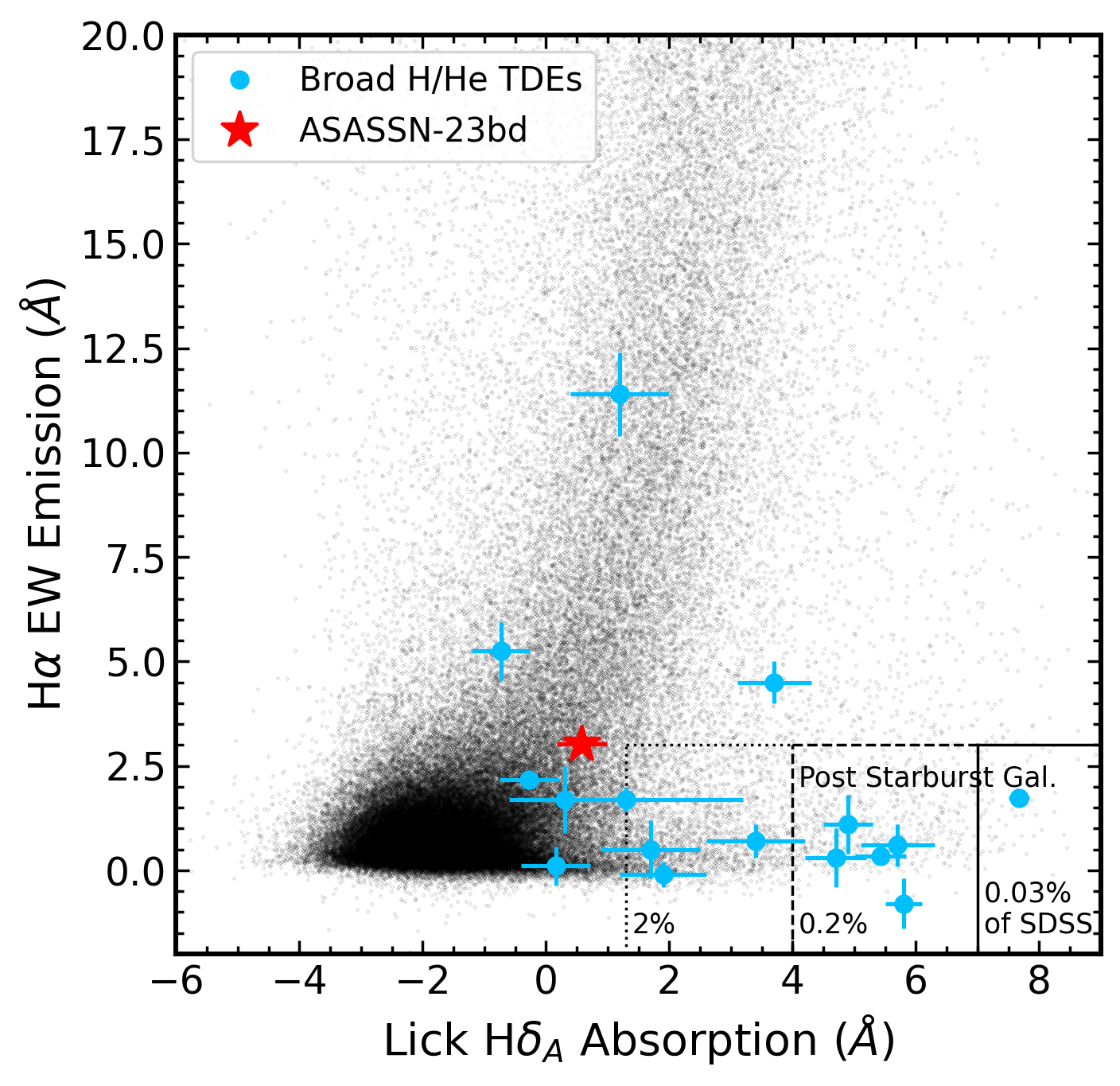} \hfill
    \includegraphics[width=0.49\textwidth]{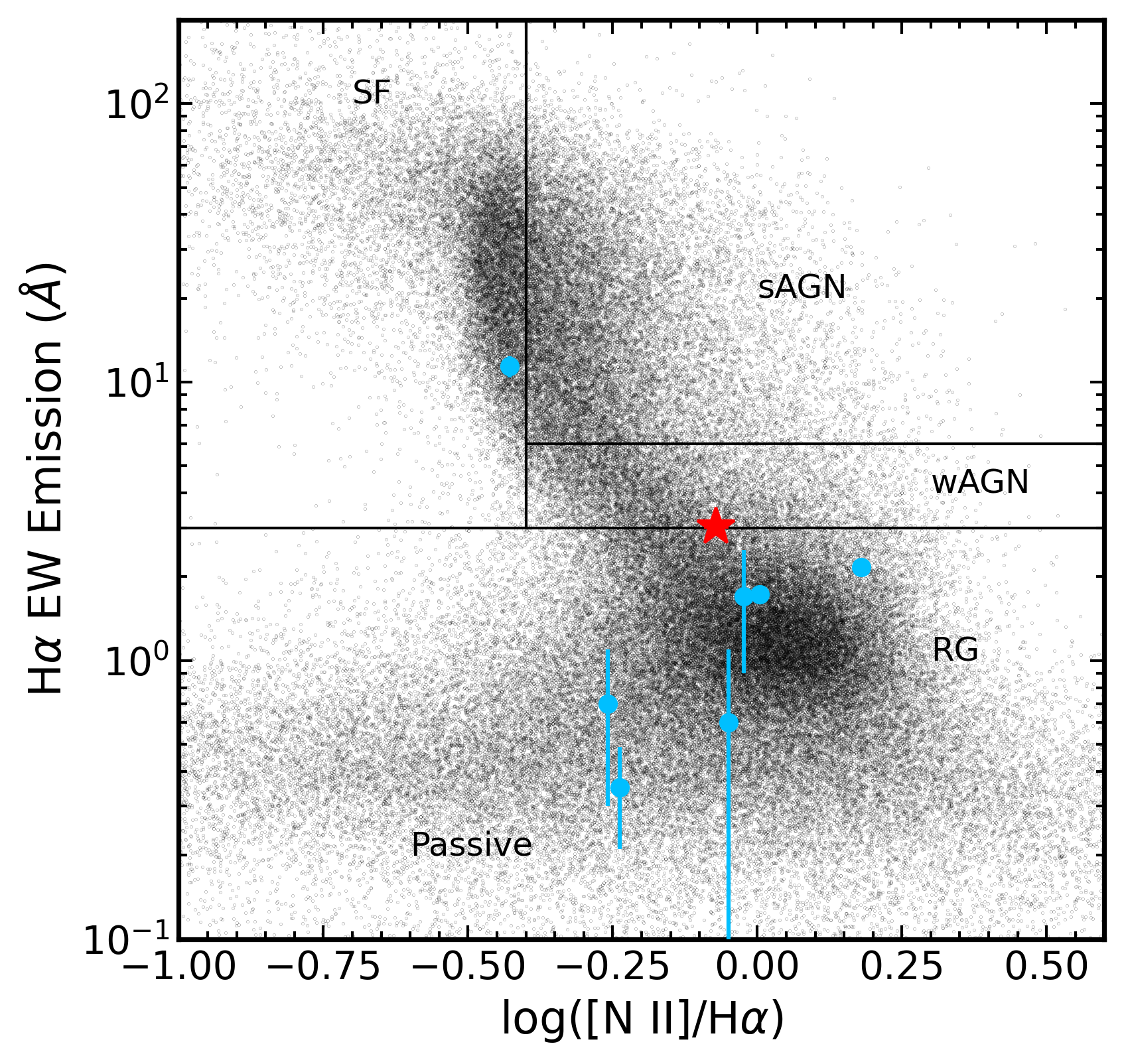} \\
    \includegraphics[width=0.49\textwidth]{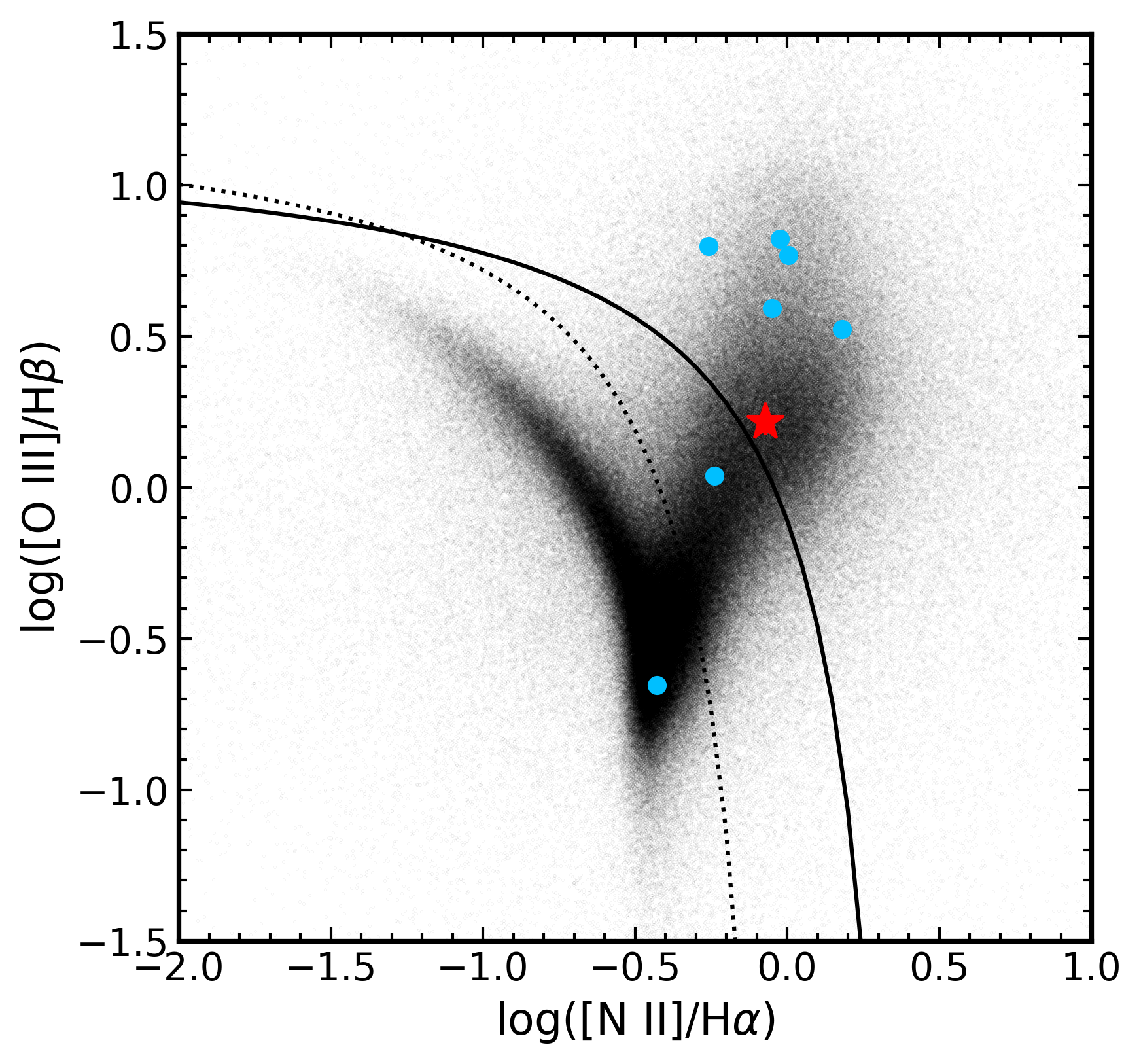} \hfill
    \includegraphics[width=0.49\textwidth]{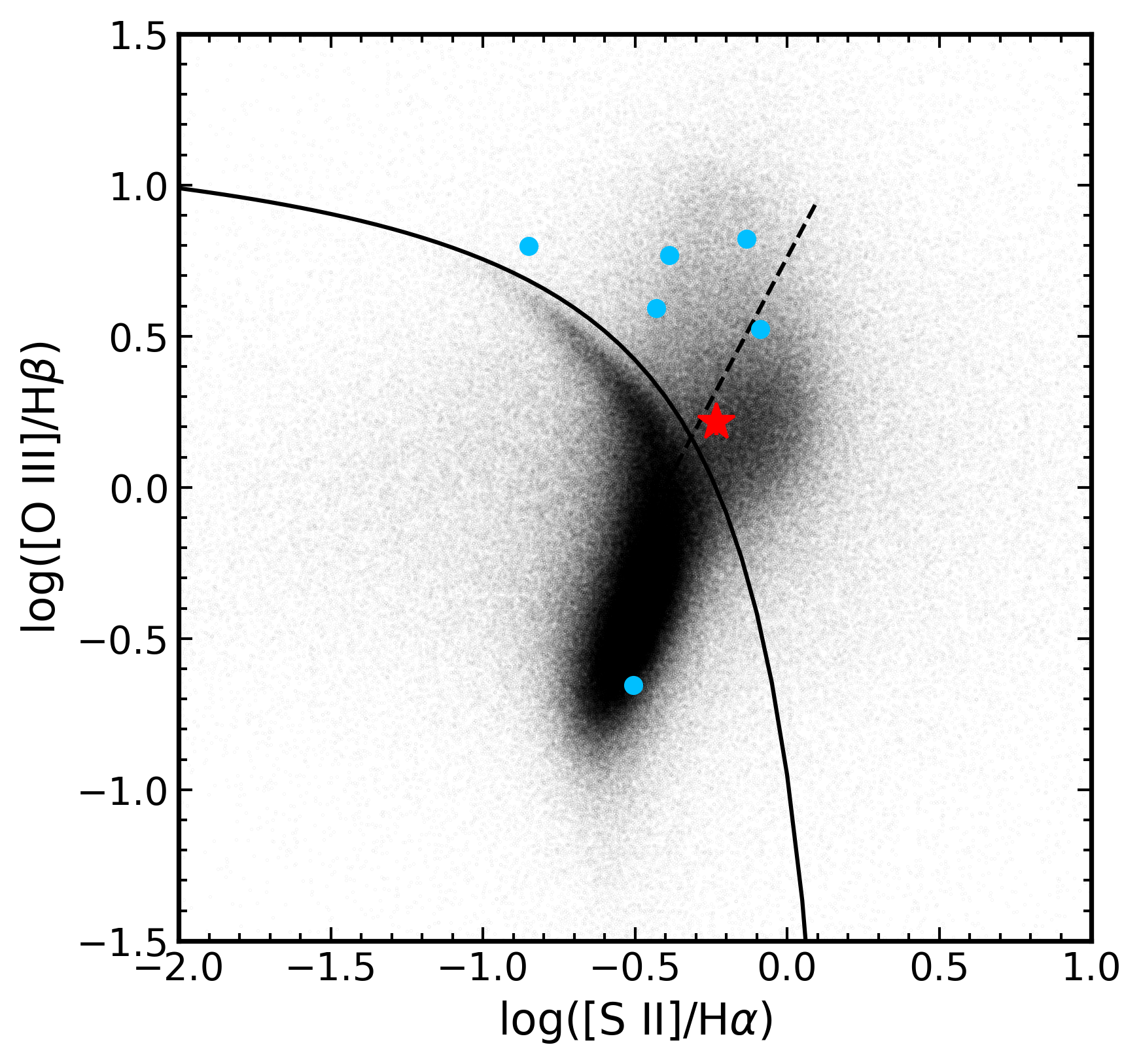} \\
    \caption{
    Host galaxy properties for selected broad H/He TDEs (blue circles) and ASASSN-23bd (red star). Black background points are galaxies from SDSS DR8 \citep{eisenstein11} in the MPA-JHU catalogue \citep{brinchmann04}.
    \emph{Upper Left Panel:}Equivalent width (EW) of the host galaxy H$\alpha$ and the Lick H$\delta_A$ absorption index. H$\alpha$ traces current star formation whereas the Lick H$\delta_A$ traces the star formation in the past Gyr. The uncertainty for NGC 3799 is the size of the marker.
    \emph{Upper Right Panel:} H$\alpha$ EW and $\log_{10}([$\NII$]/\mathrm{H}\alpha)$. Sometimes called the WHAN diagram \citep{cidfernandes11}. Plotted lines separate regions corresponding to star-forming galaxies (SF), strong AGN (sAGN), weak AGN (wAGN), and ``retired galaxies'' (RG). 
    \emph{Lower Left Panel:} $\log_{10}([$\OIII$]/\mathrm{H}\beta)$ and $\log_{10}([$\NII$]/\mathrm{H}\alpha)$ lines \citep{baldwin81,veilleux87}. The solid line separates AGNs (above the line) and \HII\ regions (below the line) \citep{kewley01}. Objects between the solid and dotted line are considered composite objects.
    \emph{Lower Right Panel:} $\log_{10}([$\OIII$]/\mathrm{H}\beta)$ and $\log_{10}([$\SII$]/\mathrm{H}\alpha)$ \citep{veilleux87}. The AGN region lies above the solid line and \HII\ regions lie below the line. The dashed line is a theoretical separation between Seyferts, located above and to the left, and LINERS, located below and to the right \citep{kewley06}.
    }
    \label{fig:galaxy}
\end{figure*}

Figure \ref{fig:galaxy} compares NGC 3799 with the host galaxies of Broad H/He TDEs and the spectroscopic properties of SDSS DR8 \citep{eisenstein11} galaxies from \citet{brinchmann04}. The top left panel shows the equivalent width of H$\alpha$ and the Lick H$\delta_A$ absorption index. H$\alpha$ is a current star-formation indicator, while the Lick H$\delta_A$ absorption index provides information about the past Gyr of star formation to identify post-starburst galaxies. The boxes are from \citet{french16} and indicate the host galaxy's evolutionary stage. TDEs generally prefer E+A or peculiar host galaxies \citep{arcavi14,french16,hammerstein21}, but because TDEs can occur in much fainter galaxies than SDSS can obtain spectra for, a complete picture of TDE host properties requires a correction for the difference in the flux limit (see, e.g., \citealp{hammerstein21}) that is not used in this work.

The top right panel shows the H$\alpha$ equivalent width and $\log_{10}([$\NII$]/\mathrm{H}\alpha)$, which helps to discriminate between ionization mechanisms, especially those of LINER galaxies \citep{cidfernandes11}. The bottom left panel of Figure \ref{fig:galaxy} shows a galaxy diagnostic based on $\log_{10}([$\OIII$]/\mathrm{H}\beta)$ and $\log_{10}([$\NII$]/\mathrm{H}\alpha)$, and the bottom right shows another one using $\log_{10}([$\OIII$]/\mathrm{H}\beta)$ and $\log_{10}([$\SII$]/\mathrm{H}\alpha)$ \citep{baldwin81,veilleux87}. The numerical values for each parameter are 
$EW_{H\alpha} = 3.03\pm0.13$ \AA, 
$EW_{H\delta_A} = 0.58\pm0.41$ \AA, 
$\log_{10}([$\NII$]/\mathrm{H}\alpha) = -0.071 \pm 0.021$, 
$\log_{10}([$\SII$]/\mathrm{H}\alpha) = -0.233 \pm 0.032$, and 
$\log_{10}([$\OIII$]/\mathrm{H}\beta) = 0.215 \pm 0.042$.
Finally, we note that NGC 3799 has a $W1-W2$ color of $-$0.03 mag, which indicates no strong AGN activity \citep{Stern12}. In summary, Figure \ref{fig:galaxy} shows NGC 3799 is still undergoing star formation and hosts weak AGN activity.

We use the photometry in Table \ref{tab:host} to estimate several host-galaxy parameters using the Fitting and Assessment of Synthetic Templates package (\verb|FAST|; \citealt{kriek09}). We assume a \citet{salpeter55} initial mass function, an exponentially declining star-formation rate, and the stellar population models of \citet{bruzual03}. This population model uses light from the entire host galaxy. However, there is evidence for an older, quiescent nucleus from the SDSS and Pan-STARRS host-galaxy images. The star formation is likely associated with the spiral arms and driven by NGC 3799's tidal interactions with its larger nearby neighbor, NGC 3800. The model results give an age of 2.0$^{+0.2}_{-0.4}$ Gyr, a stellar mass of $6.3^{+0.6}_{-0.6}\times10^{9} M_\odot$, a star formation rate of $1.2_{-0.6}^{+0.1}\times10^{-1}M_\odot$ yr$^{-1}$, and a specific star formation rate of $2.5^{+1.3}_{-0.3}\times10^{-11}$ yr$^{-1}$. For the \citet{reines15} scaling relation, the host stellar mass implies a black hole mass of $M_{BH} = (1.6\pm1.0)\times10^{6}$ M$_\odot$, which is similar to other FaF TDEs (e.g., \citealt{Charalampopoulos22} and \citealt{blagorodnova17}).

Since NGC 3799 is classified as a LINER, we searched for prior X-ray emission that could indicate possible AGN activity using archival \emph{Swift} XRT observations obtained before ASASSN-23bd. We find no evidence of prior X-ray emission from the host. Merging all available archival observations, we obtain a 3$\sigma$ upper limit to the 0.3-10.0 keV count rate of 0.002 counts s$^{-1}$. Assuming an absorbed power law with a photon index $\Gamma=1.7$ \citep[e.g.,][]{ricci17} and a Galactic column density of $2.51\times10^{20}$ cm$^{-2}$ \citep{HI4PI16}, we derive an absorbed luminosity of $< 2\times10^{40}$ erg s$^{-1}$. This limit rules out strong, but not weak, AGN activity, consistent with Figure \ref{fig:galaxy}. 

It appears that ASASSN-23bd is not significantly affected by extinction. The Milky-Way extinction towards NGC~3799 is $A_V=0.085$ mag \citep{schlafly11}. We use four different methods to estimate the host-galaxy extinction. 
First, the pseudo-equivalent width of the \NaI\ D line in the SDSS spectrum can be used to estimate an extinction using the relationship of \citet{Poznanski12}. We perform a 1500 iteration bootstrap on the \NaI\ D line and take the median of the 1500 resulting extinction values to derive $A_V= 0.55\pm0.09$ mag. 
Second, the \verb|FAST| fit provides an extinction estimate of $A_V = 0.09_{-0.01}^{+0.36}$ mag. 
Third, using the SDSS line fluxes, the Balmer decrement is $3.7\pm0.3$. Using the Balmer decrement - extinction relationship from \citet{Dominguez13}, we derive $E(B-V) = 0.22$~mag and $A_V = 0.88$~mag, although with $R_V = 3.1$, this drops to $A_V=0.66$~mag.
However, these estimates may be inaccurate for several reasons. First, they are derived using emission from the larger narrow-line region and thus may overestimate the line-of-sight extinction toward the nucleus. Second, the \verb|FAST| population model integrates emission from the whole galaxy, but there is evidence for an older, quiescent nucleus with star formation associated with the spiral arms and possible tidal interactions with its larger nearby neighbor, NGC 3800. Third, the Balmer decrement assumes Case B recombination as the only source of H-line production. But NGC 3799 is a LINER, implying that shocks may modify the line ratios, flattening the Balmer decrement and reducing the extinction estimate. This is difficult to disentangle from the effects of dust. There is no significant \NaI\ D line in our follow-up spectra where the nuclear transient dominates the emission, although \citet{Charalampopoulos24} report \NaI\ D in their spectroscopic data. However, we find the upper limit on \NaI\ D absorption from the LRIS spectrum corresponds to $A_V \le 0.05$ mag. This is a line-of-site estimate directly from the TDE observations, so we assume that the host-galaxy extinction of ASASSN-23bd is negligible. 

\begin{table}
\centering
\caption{Archival photometry of NGC 3799 from Swift (UVW2, UVM2, UVW1; \citealt{gehrels04,poole08,Breeveld11}), SDSS ($u$, $g$, $r$, $i$, $z$; \citealt{aguado19}), 2MASS (J, H, K$_s$; \citealt{skrutskie06}), and WISE (W1 and W2; \citealt{wright10}). All photometry is in the AB system.}\label{tab:host}
    \begin{tabular}{ccc}
    Filter & Magnitude & Magnitude Uncertainty\\
    \hline
    $UVW2$    &   16.57   &   0.02    \\
    $UVM2$    &   16.89   &   0.03    \\
    $UVW1$    &   16.31   &   0.02    \\
    $u$       &   15.60   &   0.01    \\
    $g$       &   14.23   &   0.01    \\
    $r$       &   13.71   &   0.01    \\
    $i$       &   13.41   &   0.01    \\
    $z$       &   13.25   &   0.01    \\
    $J$       &   12.90   &   0.03    \\
    $H$       &   12.80   &   0.04    \\
    $K_s$     &   12.95   &   0.04    \\
    $W1$      &   14.27   &   0.02    \\
    $W2$      &   14.94   &   0.02    \\
    \hline
    \end{tabular}
\end{table}

\begin{figure*}
\includegraphics[width=\linewidth]{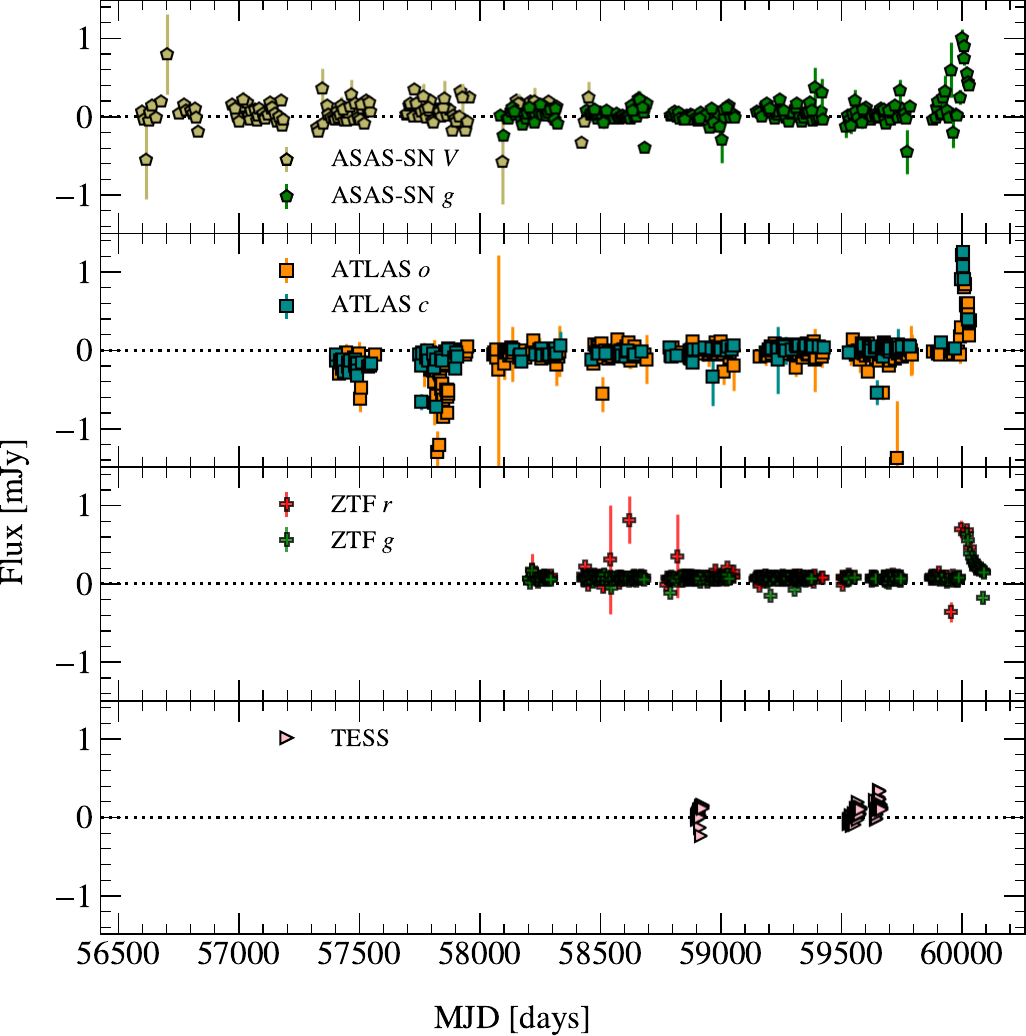}
\caption{Stacked archival flux light curves for NGC 3799 with 3$\sigma$ uncertainties. There are no prior AGN flares detected in NGC 3799. The ASAS-SN $V$- and $g$-band data are the khaki and green pentagons, respectively; the ZTF $g$- and $r$-band data are the green and red plusses, respectively; the TESS data are pink triangles; and the ATLAS $c$- and $o$-band data are cyan orange squares, respectively. The ATLAS data near MJD 57800 pass reasonable ATLAS quality cuts but may nonetheless be adversely affected from clouds.
\label{fig:AGN}}
\end{figure*}

Since NGC 3799 hosts weak AGN activity, we searched the archival ASAS-SN, ATLAS, ZTF, and TESS observations for AGN variability or outbursts prior to ASASSN-23bd. The forced-photometry light curves are shown in Figure \ref{fig:AGN} and show no evidence of variability. We find that the RMS is $\le100$ $\mu$Jy in each TESS sector, with sectors 22 and 49 having the largest RMS at $90$ $\mu$Jy. The ATLAS $o$ band has the largest RMS variance out of all the survey data with a value of $263$ $\mu$Jy, corresponding to a luminosity of $\nu L_\nu = 3.6\times10^{41}$ erg s$^{-1}$. If this were real and corresponded to a 1\% variability amplitude, then the AGN has $\nu L_\nu < 3.6\times10^{43}$ erg s$^{-1}$. 
With such low variance, we rule out strong AGN activity with $\nu L_\nu > 10^{41}$ erg s$^{-1}$ over the past decade.

\subsection{Photometric Analysis}\label{sec:lcs}

\begin{figure*}
\includegraphics[width=\textwidth]{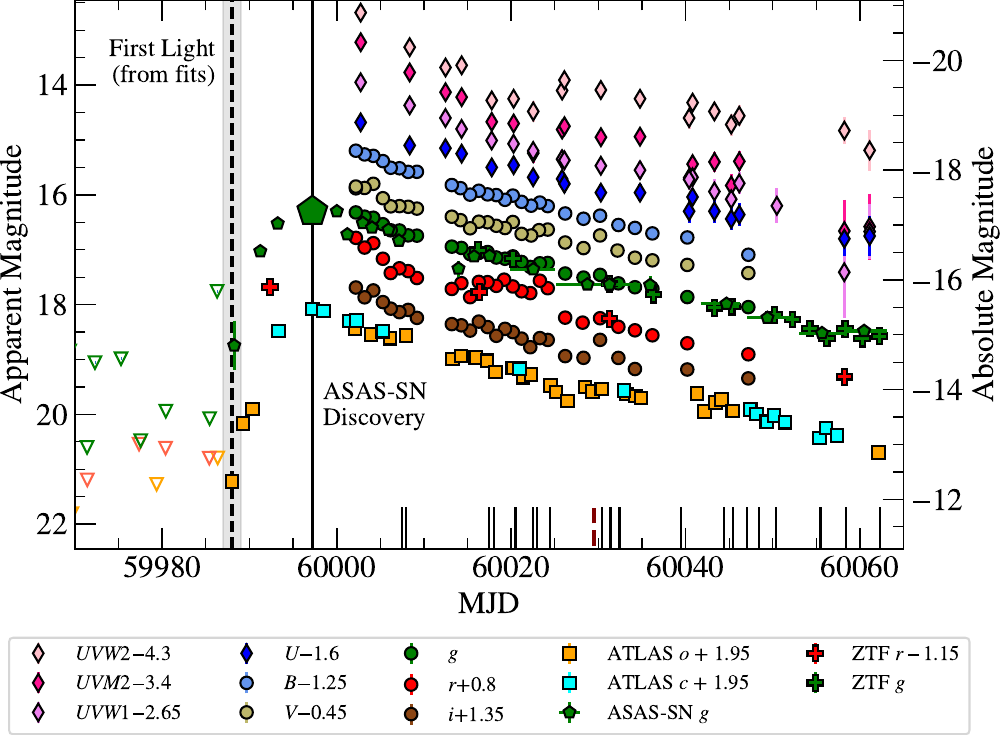}
\caption{Ultraviolet and optical light curves of ASASSN-23bd from \emph{Swift} (diamonds), Swope (circles), ZTF (pluses), ATLAS (squares), and ASAS-SN (pentagons). The ASAS-SN discovery point is denoted with an enlarged pentagonal marker and vertical black line, and the epoch of first light with a dashed black line with the grey area denoting the $3\sigma$ uncertainty in the fit that determined the epoch of first light (the one- and two-component power laws have a consistent value of $t_0$). Short, black lines along the horizontal axis correspond to optical spectral observations, along with a dotted maroon line for our NIR spectral observation. Downward-pointing triangles denote upper limits. All magnitudes are in the AB system and are host subtracted. No correction for the (small) Milky-Way extinction is applied. All points without visible uncertainties have uncertainties smaller than the data markers.
\label{fig:lc}} 
\end{figure*}

\begin{figure}
\includegraphics[width=\linewidth]{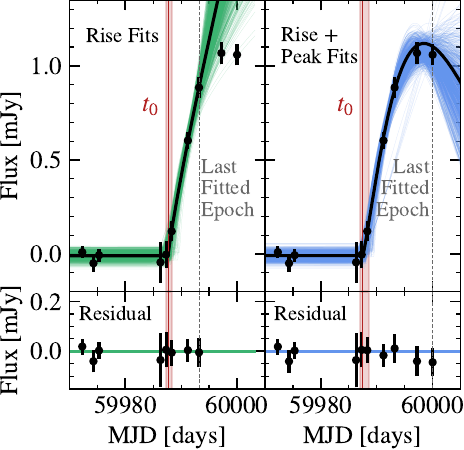}
\caption{
Single (left) and double (right) power law fits to the rise of ASASSN-23bd. The best fit is shown in black, with individual samples shown in green (left) and blue (right). The bottom panels show the fit residuals. Section \ref{sec:lcs} describes the fit models, and Table \ref{tab:mcmc} presents the fit parameters for both models.
\label{fig:risefit}
}
\end{figure}

Figure \ref{fig:lc} presents our photometric data for ASASSN-23bd. To constrain the time of peak, we perform MCMC fits with a simple parabolic model to the ASAS-SN data near peak to find $t_{\rm peak} (\mathrm{MJD})=60000^{+3}_{-3}$. 
\citet{Yao23} fit the rising light curves of 33 TDEs and find that a power-law fit is preferential to a Gaussian fit. Thus, we fit the rising ASAS-SN $g$-band photometry with two different power-law models of the form 
\begin{equation}
    f(t) = 
    \begin{cases} 
        f_0 & t < t_0 \\
        f_0+k\left(\frac{t-t_0}{1+z}\right)^a & t \geq t_0. 
    \end{cases}\label{eq:1pl}
\end{equation}
\noindent where 
\begin{align}
    k &\equiv \frac{h}{(1+z)^2}.
\end{align} 
\noindent and $\alpha$ is either a constant, single-power model as in \citet{vallely21} and \citet{hinkle21a} or 
\begin{equation}
    \alpha \equiv \alpha_1\left(1+\frac{\alpha_2(t-t_0)}{(1+z)}\right)
\end{equation}
\noindent as also used in \citet{vallely21}. While the value of $\alpha$ in the one-component model is sensitive to how much of the rising light curve is fit \citep[e.g.,][]{vallely19,vallely21}, the $\alpha_1$ parameter from the two-component model is not. We use the \verb|emcee| fitting package \citep{emcee} with the results given in Table \ref{tab:mcmc} and shown in Figure \ref{fig:risefit}. The two models produce consistent results for $f_0$, $h$, $t_0$, and $\alpha$/$\alpha_1$. In both models, the power law slope of the rise is closer to linear in time than quadratic. The two-component model can also be used to determine the peak $g$-band magnitude. The median peak value is $1.09\pm0.09$ mJy, corresponding to an AB magnitude of $16.30\pm0.03$.

\begin{table}
\centering
\caption{MCMC fit results for the light curve rise. }\label{tab:mcmc}
    \begin{tabular}{ccc}
    \hline
    Parameter & One-Component Value & Two-component Value\\
    \hline
    $f_0$ [$\mu$Jy]         &   $-1^{+21}_{-22}$      &   $1^{+22}_{-21}$       \\
    $t_0$ [MJD]             &   $59988^{+1}_{-1}$     &   $59988^{+1}_{-1}$     \\
    $h$   [$\mu$Jy]         &   $132^{+95}_{-69}$     &   $163^{+141}_{-81}$    \\
    $\alpha$/$\alpha_1$     &   $1.1^{+0.3}_{-0.3}$   &   $1.2^{+0.4}_{-0.3}$   \\
    $\alpha_2$              &                         &   $-0.1^{+0.1}_{-0.1}$  \\
    \hline
    \end{tabular}
\end{table}

To estimate the theoretical properties of ASASSN-23bd, we fit our host-subtracted light curves using the Modular Open Source Fitter for Transients package (\verb|MOSFiT|; \citealt{Guillochon17,mockler19}). \verb|MOSFiT| provides estimates of the physical properties of the star, the SMBH, and the encounter between the star and SMBH by generating bolometric light curves from predefined TDE models, subsequently deriving light curves for each photometric band, and finally fitting the derived light curves to the observed ones. The \verb|MOSFiT| results are tabulated in Table \ref{tab:MOSFiT}. The SMBH mass from these fits agrees well with the mass estimate from the host-galaxy scaling relationships presented in Section \ref{sec:host}. 

\begin{table}
\centering
\caption{MOSFiT results for ASASSN-23bd. The middle columns present the lower uncertainty, median value, and upper uncertainty for a 1-$\sigma$ confidence interval. We include only the statistical error in our reported values. For a discussion of systematic errors, see \citet{mockler19}. } \label{tab:MOSFiT}
    \begin{tabular}{ccc}
    Quantity & Value & Units\\
    \hline
    $\log R_{\rm ph0}$      & $1.2_{-0.2}^{+0.6}$  & --          \\
    $\log T_{\rm visc}$     & $0.7_{-0.2}^{+0.2}$  & d           \\
    $b$                 & $1.0_{-0.1}^{+0.1}$  & --          \\
    $\log M_h$          & $6.4_{-0.1}^{+0.1}$  & M$_\odot$   \\
    $\log\epsilon$      & $-3.2_{-0.5}^{+0.7}$ & --          \\
    $l$                 & $1.2_{-0.2}^{+0.3} $ & --          \\
    $\log n_{\rm H,host}$   & $21.3_{-0.1}^{+0.1}$ & cm$^{-2}$   \\
    $M_{\star}$         & $0.5_{-0.1}^{+0.9} $ & M$_\odot$   \\
    $t_{\rm exp}$           & $-0.3_{-0.4}^{+0.2}$ & d         \\
    $\log\sigma$        & $0.6_{-0.1}^{+0.1} $ & --          \\
    \hline
    \end{tabular} \\
    {$\log R_{\rm ph0}$: photosphere power-law constant; $\log T_{\rm visc}$: viscous delay time scale; $b$: scaled impact parameter $\beta$; $\log M_h$: SMBH mass; $\log\epsilon$: efficiency; $l$: photosphere power-law exponent; $\log n_{\rm H,host}$: local hydrogen column density; $M_{\star}$: stellar mass; $t_{exp}$: time of disruption; $\log\sigma$: model variance.}
\end{table}

\begin{figure}
\includegraphics[width=\linewidth]{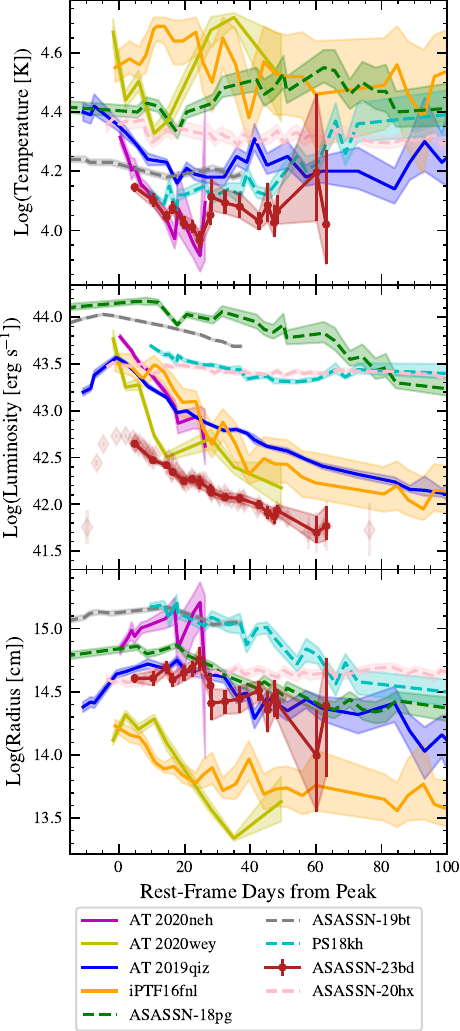}
\caption{Evolution of the blackbody temperature (top), luminosity (centre), and radius (bottom) of ASASSN-23bd (red with points) and other TDEs (other colours). FaF TDEs are shown as solid lines, whereas other nuclear transients use dashed lines. In the middle panel, the bolometrically-corrected ASAS-SN photometry is shown in faint, red diamonds, assuming only a Milky Way extinction and the blackbody fits described in the text. The bolometric luminosity from \emph{Swift} is shown in bold, red circles.
\label{fig:BB}}
\end{figure}

\begin{figure}
\includegraphics[width=\linewidth]{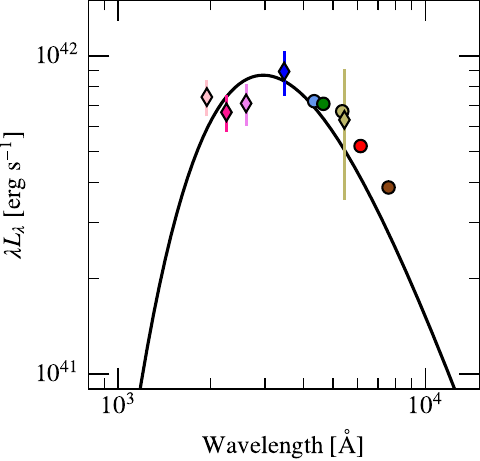}
\caption{The best-fitting UV/optical blackbody (black line) of ASASSN-23bd at MJD $\sim$60030. The photometric data have the same markers and colors as Figure \ref{fig:lc}.
\label{fig:SED}}
\end{figure}

We fit a blackbody model to the optical and UV data. Figure \ref{fig:BB} shows the resulting temperature, luminosity, and radius, and Figure \ref{fig:SED} shows the best-fit blackbody at the epoch of the NIR spectrum. We only correct for the effects of the Milky Way extinction and do not include the host-galaxy extinction since it appears to be negligible (although \citealt{Charalampopoulos24} argue for a significant host-galaxy extinction component). The results show a nearly constant blackbody temperature and radius and a decreasing blackbody luminosity with time. Our blackbody fits are consistent with the fits performed by \citet{Zhu23} using \verb|SUPERBOL| \citep{Nicholl18}.

\begin{figure}
\includegraphics[width=\linewidth]{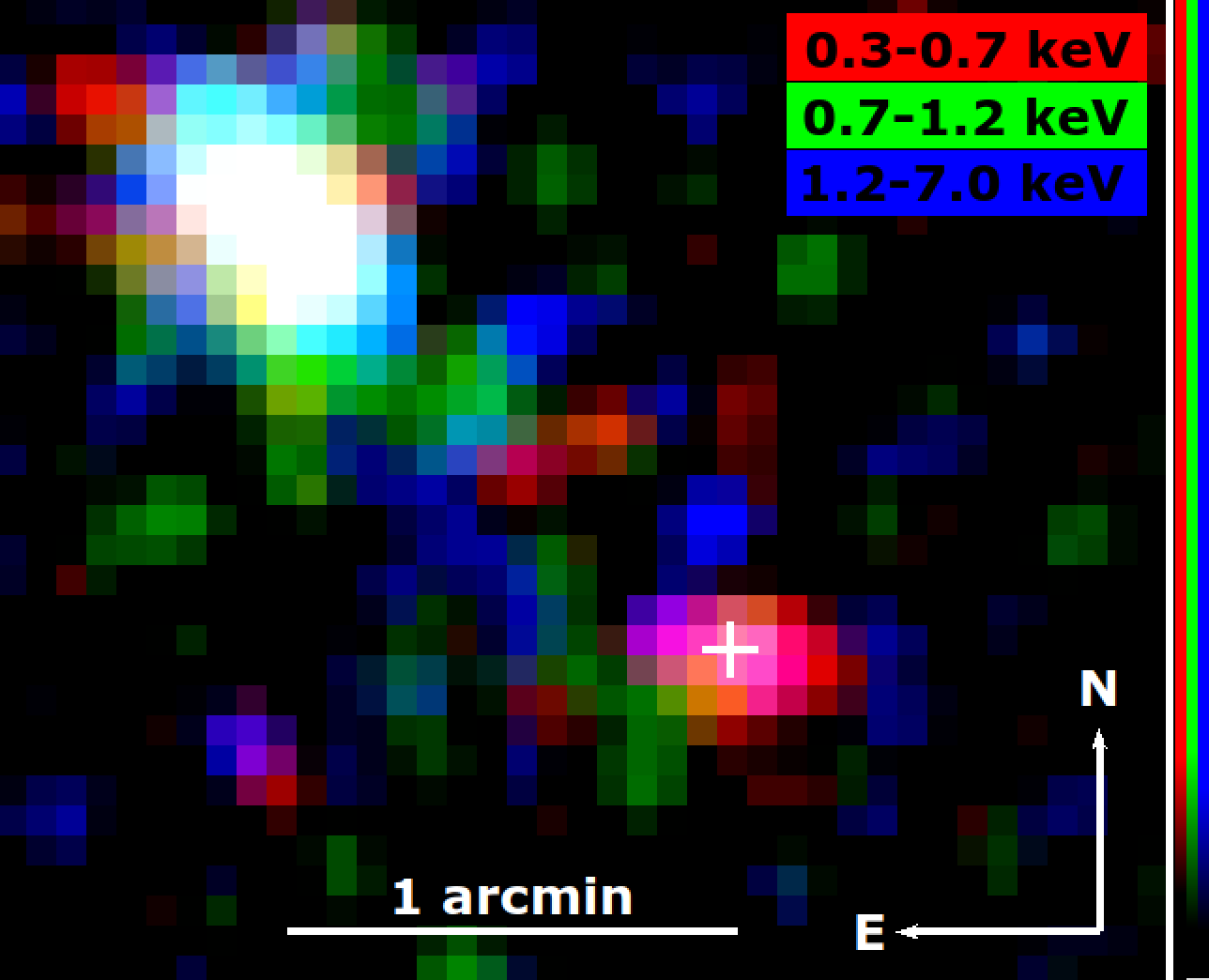}
\caption{XMM-Newton 3-color image. The image is smoothed with a 2-pixel Gaussian. The extended emission to the northeast is NGC 3800, the larger, interacting neighbor of NGC 3799. The red, green, and blue colours correspond to 300-700 eV, 700-1200 eV, and 1200-7000 eV, respectively. The white plus sign is the location of ASASSN-23bd.
\label{fig:XMM}}
\end{figure}

Figure \ref{fig:XMM} shows a smoothed, 3-color image of the \emph{XMM-Newton} detection created using the standard pipeline bands for \textsc{etruecolor}\footnote{\href{https://heasarc.gsfc.nasa.gov/docs/xmm/sas/USG/etruecolor.html}{https://heasarc.gsfc.nasa.gov/docs/xmm/sas/USG/etruecolor.html}}. The \emph{XMM-Newton} data implies a blackbody with $L\approx4\times10^{39}$ erg s$^{-1}$ and $kT = 0.1$ keV, and thus, a (spherical) photospheric radius of $\sim$1.$8\times10^{9}$~cm (see Section \ref{sec:XMM}), which is much smaller than the Schwarzschild radius for a 10$^6$ $M_\odot$ SMBH of $\sim$3.$0\times10^{12}$~cm. An unphysically small blackbody radius for the X-ray emission is a common occurrence in TDEs (e.g., \citealt{brown17a, hinkle21c}). \citet{Mummery21} showed that assuming a single-temperature, spherical blackbody can underestimate the TDE accretion disc size by up to an order of magnitude.

\subsection{Spectroscopic Analysis}\label{sec:specs}
\begin{figure*}
\includegraphics[width=\textwidth]{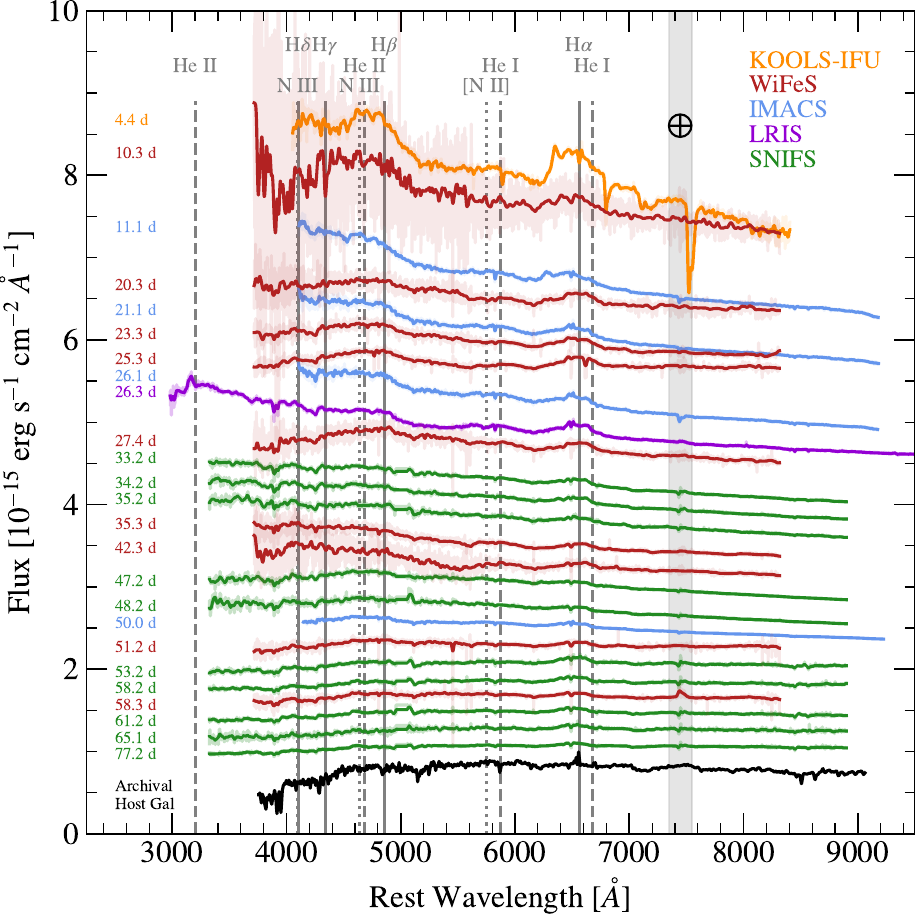}
\caption{Optical spectra of ASASSN-23bd corrected to match the observed Swope $g$, $r$, and $i$ photometry. Common H, He, and N lines are marked by solid, dashed, and dotted grey lines, respectively. The fainter lines are the original spectra, and the bold lines are smoothed spectra. Our spectra range from $+4.4$~days (top) after discovery to $+77.2$~days (bottom) after discovery. Spectra are from the KOOLS-IFU (orange), WiFeS (red), IMACS (blue), SNIFS (green), and LRIS (purple). The archival SDSS spectrum is in black. We do not have access to the original KOOLS-IFU data, so we cannot improve the data reduction. Besides the telluric feature, which is redder than the rest of our spectra, the wavelengths align with the rest of our data.
\label{fig:spectra}}
\end{figure*}

\begin{figure*}
\includegraphics[width=\linewidth]{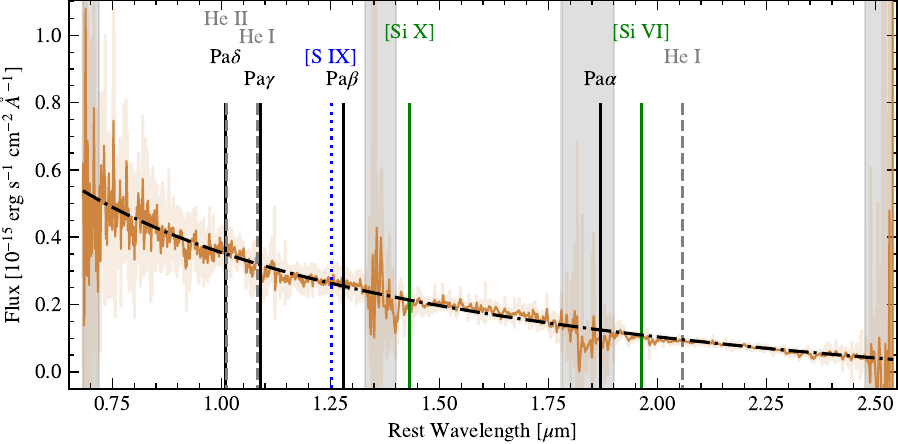}
\caption{Rest-frame NIR spectrum of ASASSN-23bd. No AGN features, such as H, He, or coronal emission lines, are visible in the spectrum. Solid grey lines are Paschen series H lines; the dashed grey line is \HeI $\lambda$ 1.083. Coronal S and Si lines are denoted with dotted blue and solid green lines, respectively. The middle two grey regions denote regions contaminated by telluric absorption. The shaded grey areas are data excluded from the fit; the dash-dot black line is the best power-law fit.
\label{fig:nir}}
\end{figure*}

The optical spectra of ASASSN-23bd are presented in Figure \ref{fig:spectra}. We photometrically calibrated (also known as color matching or ``mangling''; \citealt{hsiao07}) the spectra using the optical photometry from POISE. Synthetic fluxes were calculated from each spectrum and compared to the POISE fluxes. The flux ratios are fit by a spline, which is then used to correct the spectrum. This process is repeated until the synthetic fluxes agree with the POISE fluxes. Our NIR spectrum is presented in Figure \ref{fig:nir} where no mangling was applied. ASASSN-23bd lacks the NIR spectral features commonly observed in AGNs such as H, He, or coronal emission lines. The lack of coronal lines seems to require having little nuclear gas. We fit a power law of the form $F_\lambda = a\times\lambda^b+c$ to our NIR spectrum. The best fit values are $a=4.2\times10^{-14}$~erg s$^{-1}$ cm$^{-2}$ \AA$^{-1}$, $b=-0.54$, and $c=-4.5\times10^{-16}$~erg s$^{-1}$ cm$^{-2}$ \AA$^{-1}$. However, because the NIR is not host subtracted, there is most likely host contamination, which may influence the fitted power-law slope.

\begin{figure*}
\includegraphics[width=\textwidth]{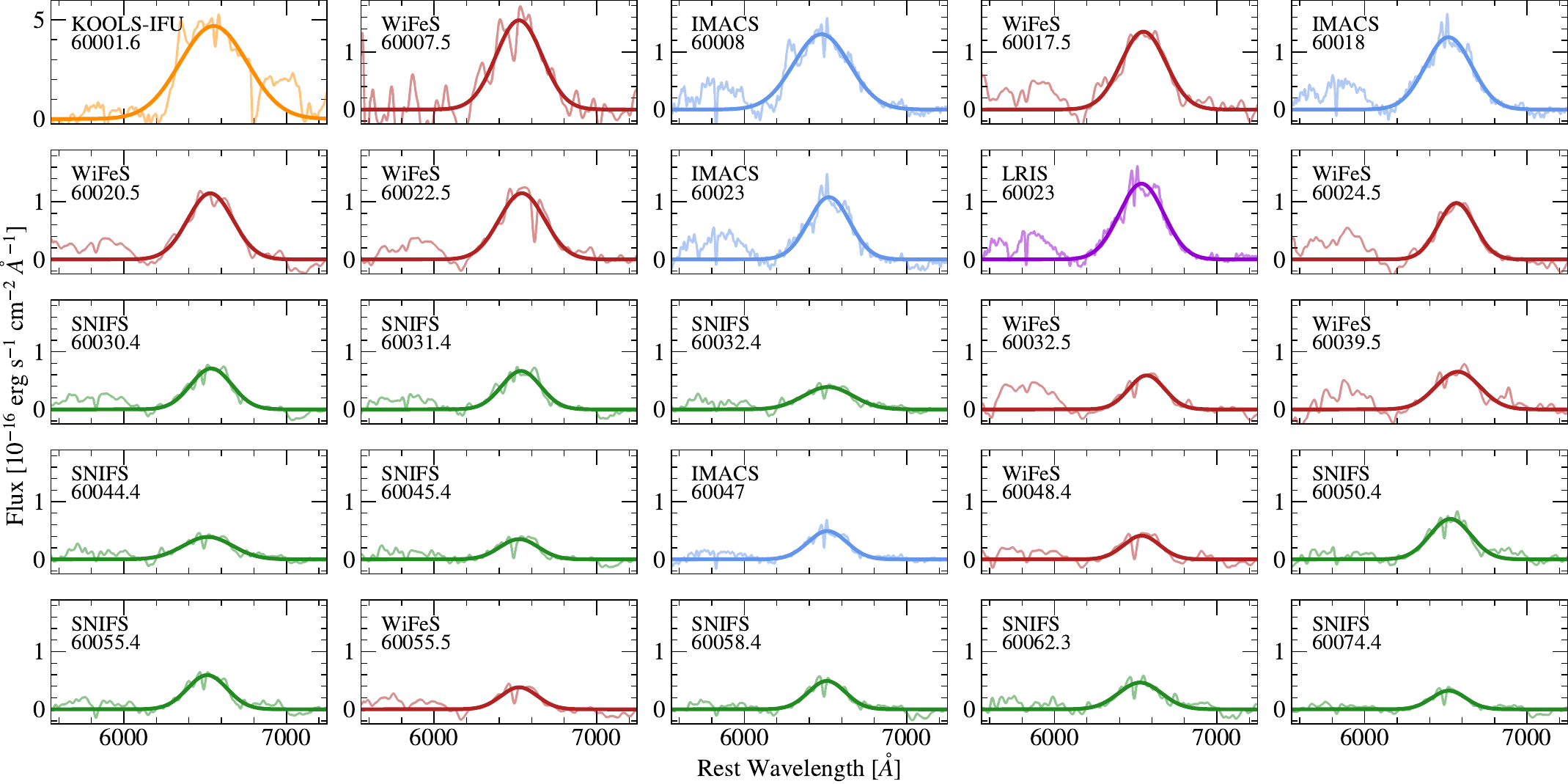}
\caption{H$\alpha$ emission features and their fitted Gaussian profiles. The color scheme is the same as in Figure \ref{fig:spectra}. The spectra were continuum subtracted using a linear fit to each spectrum between $6100$ \AA\ $\le \lambda \le 6230$ \AA\ and  $7600$ \AA\ $\le \lambda \le 7900$ \AA. A smoothing function is applied to the plotted spectra.
\label{fig:Halpha}}
\end{figure*}

\begin{figure}
\includegraphics[width=\linewidth]{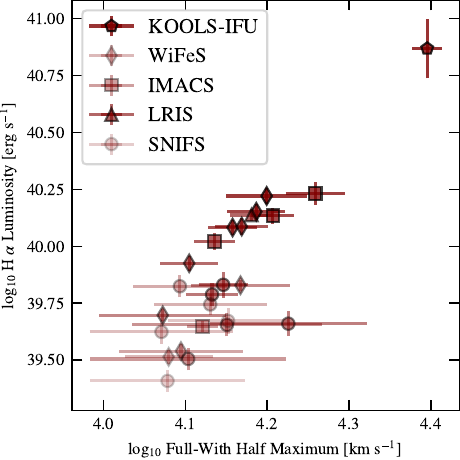}
\caption{The H$\alpha$ luminosities and FWHM of ASASSN-23bd. There is a weak linear trend between the width and luminosity of the feature, with wider features generally being more luminous. Each point is shaded based on epoch, with earlier epochs being darker and later epochs being lighter. Different symbols are used based on the instrument used to acquire the data.
\label{fig:lum-fwhm}}
\end{figure}

We see prominent H$\alpha$ and weak H$\delta$ and \HeI\ emission in all our optical spectra, along with \HeII\ $\lambda$4686 in the earliest spectra. While there is also \HeII\ $\lambda$3100 emission in the Keck-I LRIS spectrum. The rest of our spectral data do not extend this far into the blue. From  these features, ASASSN-23bd is a member of the H+He TDE class \citep{leloudas19,vanvelzen21a}.
We fit the H$\alpha$ line with a Gaussian using a linear continuum model normalized to the line bracketing regions $6100$ \AA\ $\le \lambda \le 6230$ \AA\ and  $7600$ \AA\ $\le \lambda \le 7900$ \AA. Figure \ref{fig:Halpha} shows the evolution of the H$\alpha$ feature over time, as well as our Gaussian fits. Figure \ref{fig:lum-fwhm} displays the integrated luminosity of each Gaussian and its FWHM. As the phase increases, the H$\alpha$ luminosity monotonically declines. We find evidence for a correlation between H$\alpha$ integrated luminosity and FWHM with a Kendall $\tau$ value of 0.473 corresponding to a $p$ value of 0.001, suggesting a weak, linear trend. This is the typical evolution of a TDE, where spectral lines narrow as they become less luminous.

\section{Discussion}\label{sec:discussion}

\subsection{TDE or AGN Flare?}\label{sec:is_flare?}
While ASASSN-23bd was initially classified as a TDE and \citet{Zhu23} rule out a Type II supernova due to the lack of low-ionization metals in the spectra near peak, the indicators of weak AGN activity in NGC 3799 might imply that ASASSN-23bd could instead be an AGN flare. For example, ``starved”, low-luminosity AGNs such as NGC 3799 may have occasional flares that resemble TDEs \citep{Saxton18}. In this section, we evaluate the observations of ASASSN-23bd to determine if it is indeed a TDE or, instead, an AGN flare from an otherwise quiescent AGN.

We observe a rapid rise to peak and a smooth decline thereafter, a hot, constant blackbody temperature, no \MgII\ or \FeII\ emission features, and a central SMBH mass less than $10^8M_\odot$. These characteristics are generally attributed to TDEs rather than AGN flares \citep{Zabludoff21,frederick21}. Our NIR spectrum does not show spectral features, unlike the spectra of AGN and CL-AGN \citep{Landt08, Neustadt23}. Furthermore, a LINER designation does not necessarily signify the presence of an AGN. The line ratios seen for LINERs may also originate from post-AGB stars \citep{Yan12} or merger-induced shocks \citep{Rich15}.

The supersoft X-ray spectrum detected by XMM-Newton $\sim$95 days post-peak also favors the TDE hypothesis. 
A supersoft spectrum would be unusual for a low-luminosity AGN experiencing stochastic variability \citep{auchettl18}. Soft X-ray variability due to a change in obscuration is typically accompanied by strong hard X-ray emission (e.g., \citealp{Sazonov07}), which we do not see here. Conversely, an observed soft X-ray spectrum is expected and commonly observed in TDEs (e.g., \citealp{auchettl17, auchettl18,  vanvelzen21a, Guolo23}). For a TDE, the earliest \emph{Swift} non-detection implies that any earlier X-ray peak must be very weak or that there is only a late-time rise, as has been observed in other TDEs (e.g., ASASSN-15oi \citealp{gezari17, holoien18a}, OGLE16aaa \citealp{Kajava20}, ASASSN-19dj \citealp{hinkle21a}, and eRASSt J074426.3+291606 \citealp{malyali23}). While the X-ray data do not conclusively point to a TDE interpretation, a growing number of AGN also show soft X-ray emission (e.g., \citealp{Boller21, Jiang22, Sacchi23}), the combined X-ray, UV, optical, and NIR evidence argues for a TDE interpretation.

\subsection{The Early-Time Rises of TDEs and Other Nuclear Transients}\label{sec:RISES_DISCUSSION}
There are early-time light curve fits for the TDEs ASASSN-19bt \citep{holoien19b}, ASASSN-19dj \citep{hinkle21a}, ZTF19abzrhgq/AT~2019qiz \citep{nicholl20}, ZTF20acitpfz/AT~2020wey \citep{Charalampopoulos23}, and ASASSN-22ci/AT~2022dbl (Hinkle et al. 2024, in prep), all of which are consistent with a rise of $f \propto t^2$. ASASSN-23bd, however, has a nearly linear rise. ASASSN-23bd is not alone in this regard: other nuclear transients such as the repeating partial TDE ASASSN-14ko \citep{payne21, Tucker21, Payne22, Cufari22, Liu23} and the ANT ASASSN-20hx \citep{hinkle22a} also display linear rises.  

Interestingly, the majority of TDEs with $t^2$ rises have hosts that do not display AGN activity, whereas the hosts of the linearly-rising ASASSN-14ko, ASASSN-20hx, and ASASSN-23bd show signs of AGN activity. Thus, AGN activity and early-time light curve rise shape may be connected since TDEs in host galaxies with AGN activity predominantly have $f \propto t$ rises while those in non-AGN hosts exclusively have $f\propto t^2$ rises. There are currently no theoretical explanations for the trends seen in the early-time rises of TDEs, including the now-apparent bimodality between linear and quadratic rises. There are no models for the origin of a $f \propto t$ rise as opposed to a $f \propto t^2$ rise, which can be caused by a constant expansion velocity at a fixed photospheric temperature. Further early-time observations of TDEs are needed to determine if there is a correlation.

\subsection{Comparison to Other Low-Redshift Nuclear Transients}
While ASASSN-23bd is the lowest-luminosity optically-selected TDE to date, several other nuclear transients at low redshift have been claimed as TDEs. Many of these sources have lower UV/optical luminosities than ASASSN-23bd, either because they are intrinsically fainter at these wavelengths or are more heavily obscured than a typical TDE.

First, \citet{malyali23} present the "low luminosity and slow" TDE eRASSt J074426.3+291606 (J0744), an X-ray-selected TDE that has a low intrinsic ultraviolet/optical luminosity. Given the significant X-ray emission, it is unsurprising that J0744 has a greater bolometric luminosity than ASASSN-23bd. Unlike FaF TDEs, J0744 declines slowly. This may be explained by either photon trapping within an outflow \citep{metzger16} or low circularisation efficiency \citep{Steinberg24}). One prediction of circularisation efficiency variations is delayed soft X-ray emission in FaF TDEs, so future X-ray follow-up of ASASSN-23bd will help validate the mechanism differentiating between "low luminosity and slow" and FaF TDEs. 

Second, \citet{Panagiotou23} claim WTP14adbjsh, a heavily-obscured nuclear flare discovered in the IR at $z=0.0106$, is a TDE, but there is no definitive evidence for a TDE interpretation. Similar to ASASSN-23bd, the pre-flare WISE $W1 - W2$ color of the host cannot rule out weak AGN activity, nor does its location on galaxy diagnostic diagrams (e.g., the WHAN diagram) exclude LINER-like or weak AGN behaviour. While optical surveys do not find previous AGN variability, the large extinction ($A_V\approx9$~mag based on their \NaI\ D EW measurements and the relationship in \citealt{Poznanski12}) prevents robust constraints on AGN activity from archival survey data, leaving the TDE interpretation uncertain. 

Third, \citet{Nikolajuk13} and \citet{irwin15} present observations of IGR J12580+0134, a claimed X-ray-selected TDE. While the host of IGR J12580+0134 has shown AGN activity and is a known Seyfert 2 galaxy, \citet{Nikolajuk13} argue for a TDE interpretation based on the extreme hard and soft band X-ray brightness. Hosted in NGC 4845 with a redshift of 0.003663 \citep{springob05}, this would be the lowest-redshift TDE to date.

Finally, from the arguments presented in this work, we believe that ASASSN-23bd is the strongest of these nearby TDE candidates based on the comprehensive X-ray, UV, optical, and NIR follow-up observations in this work and a radio detection \citep{Sfaradi23}. Interestingly, each low-redshift TDE candidate has some evidence for potential AGN activity, including ASASSN-23bd (although we rule out an AGN interpretation in Section \ref{sec:is_flare?}). Regardless of the classification for each object, further monitoring for delayed features such as soft X-ray emission or coronal lines like those recently detected in the FaF TDE AT~2019qiz \citep{Short23}, may improve our understanding of the long-term ramifications of accretion-powered flares on SMBHs.

\subsection{Comparison to Other TDEs}
In this subsection, we compare ASASSN-23bd to other TDEs and nuclear transients. Our comparison sample consists of iPTF16fnl \citep{blagorodnova17}, PS18kh \citep{holoien19b,vanVelzen19b}, ASASSN-18pg \citep{holoien18a}, ASASSN-19bt \citep{holoien19c}, AT~2019qiz \citep{nicholl20}, AT~2020neh \citep{Angus22}, ASASSN-20hx \citep{hinkle22a}, and AT~2020wey \citep{Charalampopoulos22}. We compiled this sample by selecting FaF TDEs and several ``normal'' TDEs observed at peak. We also include the ANT ASASSN-20hx since it also has a linearly rising early-time light curve. The data are taken from the literature except for AT~2020neh and AT~2020wey, which we re-fit using the \emph{Swift} photometry from \citet{Angus22} and \citet{Charalampopoulos22}, respectively.

In Figure \ref{fig:BB}, ASASSN-23bd has the lowest temperature and luminosity in our sample. While the temperature of ASASSN-23bd is similar to AT~2020neh and PS18kh, it is cooler than the other FaF TDEs. 
In addition to being the least luminous object in our sample, ASASSN-23bd also rapidly declines in luminosity with a decline rate similar to iPTF16fnl, AT~2019qiz, and AT~2020wey but not quite as steep as AT~2020neh.
The Eddington luminosity derived for the $M_{BH}$ estimate from Section \ref{sec:host} is $2.0\times10^{44}$ erg s$^{-1}$. The peak bolometric luminosity of ASASSN-23bd is $(5.4\pm0.4)\times10^{42}$ erg s$^{-1}$, corresponding to an Eddington ratio of $L_{peak}$/$L_{Edd}$ = $2.7 \times 10^{-2}$. This is smaller than every TDE analyzed by \citet{mockler19}, where the smallest Eddington ratio was 0.11, and two orders of magnitude smaller than the typical value of $\sim$1 found by \citet{wevers19}. ASASSN-23bd is the least optical/UV luminous TDE in the sample.
Finally, the blackbody radius of ASASSN-23bd is similar to  ASASSN-18pg, AT~2019qiz, and AT~2020neh and larger than AT~2020wey and iPTF16fnl. This may be an empirical indicator for TDEs occurring in AGN hosts since ASASSN-23bd and AT~2019qiz are also in weak AGN galaxies, but the sample size is still small.

\begin{figure}
\includegraphics[width=\linewidth]{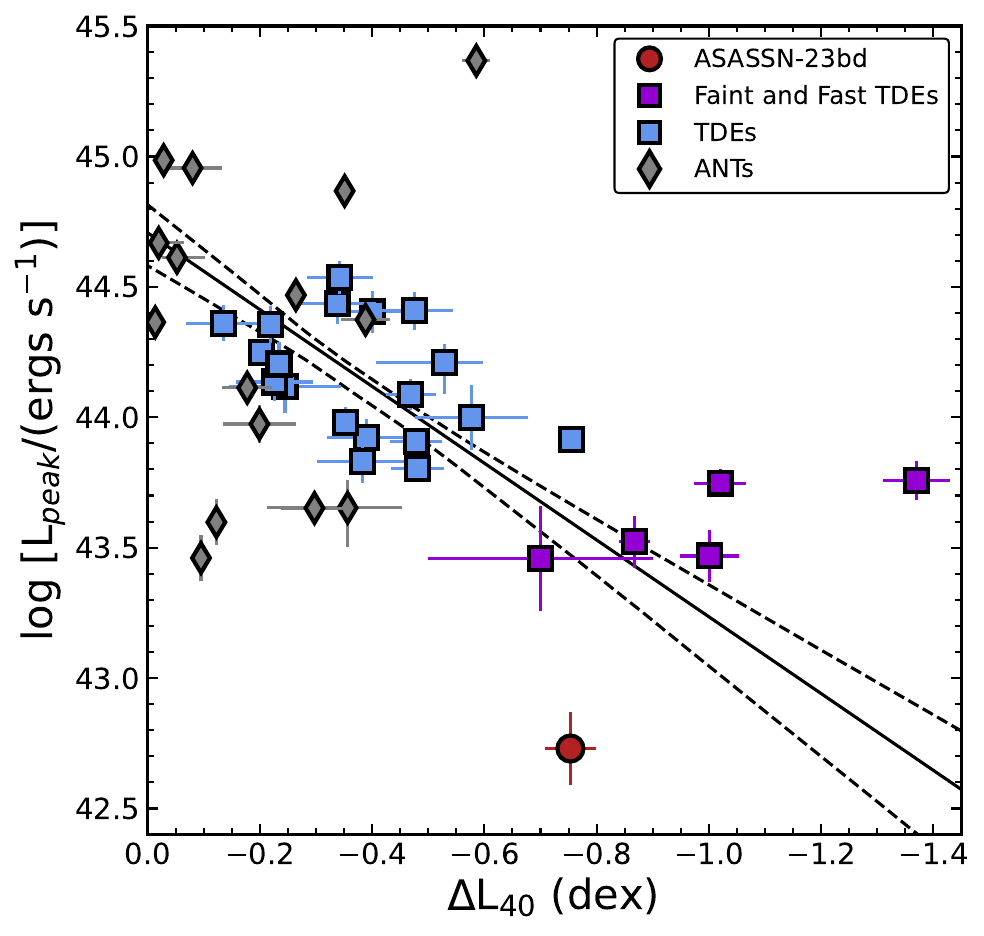}
\caption{The peak-decline relationship for TDEs \citep{hinkle20a}. ASASSN-23bd is the red circle, other FaF TDEs are the purple squares, and other slower TDEs are the blue squares. ASASSN-23bd has the lowest luminosity of our entire comparison sample. The decline of ASASSN-23bd is faster than almost all of the normal TDE in our comparison sample, but it is relatively slow compared to the other FaF TDEs. 
\label{fig:peak-decline}}
\end{figure}

Figure \ref{fig:peak-decline} shows the peak-decline relationship for TDEs \citep{hinkle20a} along with the ANTs for comparison. ASASSN-23bd is less luminous than the rest of our sample, with a peak luminosity of $\sim$10$^{42.7}$ erg s$^{-1}$, whereas the ``normal'' TDE sample clusters between $\sim$10$^{44.0}$ to $\sim$10$^{44.5}$ erg s$^{-1}$. 
The decline parameter $\Delta L_{40}$ is defined as 
\begin{equation}
    \Delta L_{40} \equiv \log_{10}\left(\frac{L_{40}}{L_{\mathrm{peak}}} \right).
\end{equation}
ASASSN-23bd has a decline parameter of $\Delta L_{40}\approx-0.7$~dex, which is faster than the normal TDEs with $\Delta L_{40}$ $\sim$$-0.2$~dex to $\sim$$-0.6$~dex, and the ANTs which have  $\Delta L_{40}$ $\sim$$0.0$~dex to $\sim$$-0.4$~dex. However, this is a slower decline than most of the other FaF TDEs.

\begin{figure}
\includegraphics[width=\linewidth]{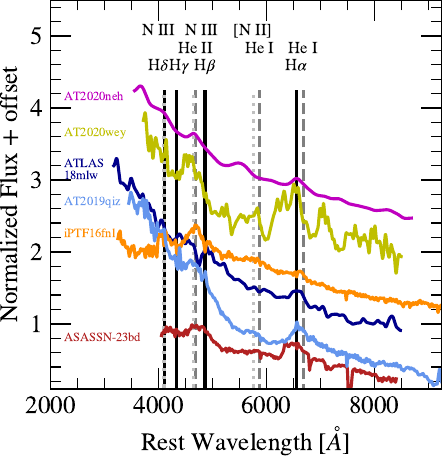}
\caption{Maximum light spectra of FaF TDEs. The flux is normalised to the median value of the entire spectrum and offset by a constant. The solid, black lines indicate H features, the dashed, grey lines indicate He features, and the dotted, grey lines indicate N features.
\label{fig:fnf_comp}}
\end{figure} 

ASASSN-23bd shows several spectral similarities to the FaF TDEs iPTF16fnl \citep{blagorodnova17}, ATLAS18mlw \citep{hinkle23a}, AT~2019qiz \citep{nicholl20}, AT~2020neh \citep{Angus22}, and AT~2020wey \citep{Charalampopoulos22}. We show the spectra closest to maximum light for each of these TDEs in Figure \ref{fig:fnf_comp}.
Both iPTF16fnl and AT~2020wey show H$\alpha$ and \HeI\ emission, whereas AT~2020neh shows only H$\alpha$. ASASSN-23bd also shows prominent H$\alpha$ emission while its \HeI\ emission is weaker than iPTF16fnl and AT~2020wey. Prior to peak, all three of these other TDEs show strong \HeII\ emission, which quickly fades post-peak. The \HeII\ line is present in our WiFes and IMACS spectra at $+4$~d and $+5$~d, respectively, and in the LRIS spectrum at $+26.3$~d.

\begin{figure}
\includegraphics[width=\linewidth]{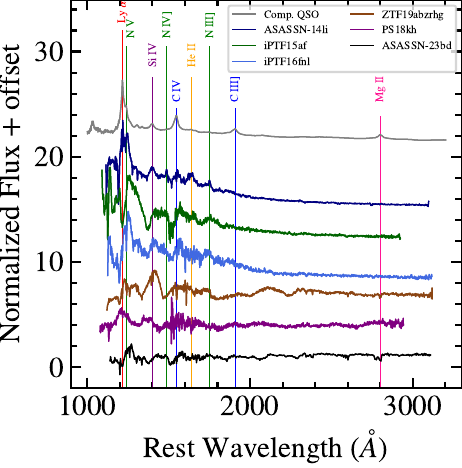}
\caption{Rest-frame UV spectrum of ASASSN-23bd and other TDEs and an SDSS composite QSO spectrum from \citep{vandenberk01}. The comparison TDEs are ASASSN-14li \citep{cenko16}, iPTF15af \citep{Blagorodnova19}, iPTF16fnl \citep{brown18}, PS18kh \citep{holoien19b,vanVelzen19b}, and ZTF19abzrhg/AT~2019qiz \citep{hung19}. The TDE spectra have been binned to $\sim$1.5\AA, normalised by their median UV flux, and offset by a constant for visibility. The phase in the caption is relative to peak, or discovery for sources not observed before the peak. Several lines commonly seen in the UV spectra of TDEs or AGNs are marked.
\label{fig:uv_spec}}
\end{figure}

Since there are limited UV spectra of TDEs, we assemble a separate comparison sample of objects with HST spectroscopy. The comparison objects are ASASSN-14li \citep{cenko16}, iPTF15af \citep{Blagorodnova19}, iPTF16fnl \citep{brown18}, PS18kh \citep{hung19}, ZTF19abzrhg/AT~2019qiz \citep{Hung21}, and the composite SDSS QSO spectrum from \citet{vandenberk01}. These TDEs and ASASSN-23bd are shown in Figure \ref{fig:uv_spec}. Compared to the other TDEs with UV spectra, ASASSN-23bd is most similar to iPTF15af \citep{Blagorodnova19} and iPTF16fnl \citep{brown18}. 
While those two objects have various emission features blueward of $\sim$1800 \AA\ and noticeable blackbody continua, the spectrum of ASASSN-23bd is much cooler and does not exhibit carbon or Lyman $\alpha$ emission lines. Conversely, all three TDEs exhibit nitrogen emission features such as N {\sc{iii}}] and N {\sc{iv}}].
This is consistent with the claim that TDEs can be enriched in nitrogen, whereas AGNs should have carbon and Mg {\sc{ii}} emission instead \citep{kochanek16a,mockler22}. Mg {\sc{ii}} appears to be a good diagnostic to differentiate between TDEs and AGNs.
No TDE to date has exhibited \MgII\ emission.

\section{Summary}\label{sec:Summary}
We present the discovery and multiband photometric and spectroscopic data of the FaF nuclear transient ASASSN-23bd in NGC 3799.  NGC 3799 is a star-forming galaxy with signs of potential weak AGN activity. ASASSN-23bd peaked on MJD $60000^{+3}_{-3}$, had a peak UV/optical luminosity of $(5.4\pm0.4)\times10^{42}$ erg s$^{-1}$, and has a near peak X-ray upper limit of $1.75\times10^{40}$ erg s$^{-1}$ and late-time X-ray detection of $L_{0.3-2 keV} = 4.2\times10^{39}$ erg $s^{-1}$. Spectroscopically, ASASSN-23bd shows $H\alpha$ emission throughout its decline, as well as \HeII\ emission near peak and \HeI\ emission shortly after the peak. The early rise of ASASSN-23bd is well fit by a near linear power law ($f\propto t^a$ with an exponent $a = 1.1\pm0.3$). We speculate that TDEs in galaxies with no AGN activity seem to rise with $f \propto t^2$ power laws, whereas TDEs in galaxies with existing AGN activity seem to prefer $f \propto t$ rises. 

The UV/optical SED of ASASSN-23bd is well fit by a blackbody, and the blackbody temperature is among the coolest of all FaF TDEs with values and evolution most similar to AT~2020neh. The luminosity is less than any other FaF TDE such as iPTF16fnl, AT~2020wey, and  AT~2020neh.

ASASSN-23bd is the lowest redshift TDE to date and due to its low luminosity, ASASSN-23bd may not have been detected if it were not so close to us. In this sense, ASASSN-23bd is a fine but fortunate addition to the collection of FaF TDEs. More FaF TDEs should be discovered by deeper sky surveys such as LSST, providing further advances in the understanding of the variety of nuclear transients.

\section*{Acknowledgements}
We thank the referee for providing helpful comments that improved this manuscript. We thank Erin Kara and Adam Malyali for their helpful comments on the manuscript. We thank Las Cumbres Observatory and its staff for their continued support of ASAS-SN. ASAS-SN is funded in part by the Gordon and Betty Moore Foundation through grants GBMF5490 and GBMF10501 to the Ohio State University and also funded in part by the Alfred P. Sloan Foundation grant G-2021-14192.

This material is based upon work supported by the National Science Foundation Graduate Research Fellowship Program under Grant No. 2236415. Any opinions, findings, and conclusions or recommendations expressed in this material are those of the author(s) and do not necessarily reflect the views of the National Science Foundation.

K.A. would also like to acknowledge Ian Price and Chris Lidman with the ANU 2.3-metre telescope.  The automation of the ANU 2.3-metre telescope was made possible through funding provided by the Centre of Gravitational Astrophysics at the Australian National University. 

C.S.K. and K.Z.S are supported by NSF grants AST-1907570 and 2307385.

M.D.S. is funded by the Independent Research Fund Denmark (IRFD, grant number  10.46540/2032-00022B )

L.G. acknowledges financial support from the Spanish Ministerio de Ciencia e Innovaci\'on (MCIN) and the Agencia Estatal de Investigaci\'on (AEI) 10.13039/501100011033 under the PID2020-115253GA-I00 HOSTFLOWS project, from Centro Superior de Investigaciones Cient\'ificas (CSIC) under the PIE project 20215AT016 and the program Unidad de Excelencia Mar\'ia de Maeztu CEX2020-001058-M, and from the Departament de Recerca i Universitats de la Generalitat de Catalunya through the 2021-SGR-01270 grant.

J.L. acknowledges support from NSF grant AAG-2206523.

This research is based on observations made with the NASA/ESA Hubble Space Telescope obtained from the Space Telescope Science Institute, which is operated by the Association of Universities for Research in Astronomy, Inc., under NASA contract NAS 5–26555. These observations are associated with program(s) GO 17001. Based on observations obtained with XMM-Newton, an ESA science mission with instruments and contributions directly funded by ESA Member States and NASA.

Based on observations gathered at the 6.5 m Magellan I (Baade) telescope and the  1 m Swope telescope at Las Campanas Observatory (Chile).

\section*{Data Availability}
 
Data is available upon reasonable request to the corresponding author. The spectra are available on WISeREP. 



\bibliographystyle{mnras}
\bibliography{asassn-23bd} 








\bsp	
\label{lastpage}
\end{document}